\documentclass[prd,aps,preprintnumbers, showpacs, nofootinbib,superscriptaddress,notitlepage]{revtex4-1}

\usepackage{amsmath,graphicx,epsfig,amssymb,dsfont,mathtools}
\usepackage[usenames]{color}
\usepackage{ulem} %% for strike-through
\usepackage{bigstrut}
\usepackage{slashed}
\usepackage{multirow}
\usepackage{subfigure}
\usepackage{xcolor}

\allowdisplaybreaks

\begin{document}

\title{Implications of recent LHCb data on CPV in b-baryon four body decays}
   \author{Qi Chen}
    \affiliation{Department of Physics and Institute of Theoretical Physics, Nanjing Normal University, Nanjing, Jiangsu 210023, China}
    \affiliation{Nanjing Key Laboratory of Particle Physics and Astrophysics}
      \author{Xin Wu}
\affiliation{Department of Physics and Institute of Theoretical Physics, Nanjing Normal University, Nanjing, Jiangsu 210023, China}
    \affiliation{Nanjing Key Laboratory of Particle Physics and Astrophysics}
\author{Zhi-Peng Xing}
\email{zpxing@nnu.edu.cn}
\affiliation{Department of Physics and Institute of Theoretical Physics, Nanjing Normal University, Nanjing, Jiangsu 210023, China}
	    \affiliation{Nanjing Key Laboratory of Particle Physics and Astrophysics}
\author{Ruilin Zhu}
\affiliation{Department of Physics and Institute of Theoretical Physics, Nanjing Normal University, Nanjing, Jiangsu 210023, China}
     \affiliation{Nanjing Key Laboratory of Particle Physics and Astrophysics}

\begin{abstract}
Motivated by the recent CPV observation, we investigate the CPV of b-baryon charmless four body decays under the U-spin symmetry. However, we find that only U-spin symmetry cannot provide effective predictions, particularly for $\Lambda_b$ decays. For giving more useful predictions, we also give a simple dynamic analysis. By counting the power($\lambda=\sqrt{\frac{\Lambda_{QCD}}{m_b}}$) of each topological diagram, we find that for the specific decay $B_b^2\to R(B_1^2  M^2 M^{\bar 2})M^2$, only one U-spin amplitude can contribute in the leading power, while for $B_b^2\to R(B_1^2  M^2)R( M^{\bar 2}M^2)$, only two U-spin amplitudes can contribute in this leading power. Then the most effective prediction can be given as
\begin{align}
    &A_{CP}^{dir}(\Lambda_b^0 \to R( p \pi^- \pi^+ )\pi^-) = (-12.99 \pm 2.83\pm2.59\pm0.65)\%,\notag
   % &A_{CP}^{dir}(\Lambda_b^0 \to R( p \pi^-)R( \pi^+ \pi^-)) = (-12.75 \pm 3.63\pm2.55)\%,\notag\\
   % &A_{CP}^{dir}(\Lambda_b^0 \to R( p \pi^-)R( K^+ K^-)) = (-65.93 \pm 20.06\pm13.90)\%,\notag\\
   % &A_{CP}^{dir}(\Lambda_b^0 \to R( p K^-)R( K^+ K^-)) = (21.28 \pm 6.08\pm4.26)\%.\notag
\end{align}
Considering the $\Lambda_b$ can effectively produced in LHCb, we strongly encourage a more precise experimental investigation of it.
\end{abstract}
\maketitle

%%%%%%%%%%%%%%%%%
\section{Introduction}
%%%%%%%%%%%%%%%%
 CP violation (CPV) is one of the most important issues in flavor physics. As it directly reflects the phase of the Cabibbo-Kobayashi-Maskawa (CKM) matrix, which is the necessary parameter in flavor-changing interactions, the study of CPV not only enables precise tests of the Standard Model (SM) but also probes new physics models. Besides, the CPV phenomenon constitutes an indispensable precondition for generating the matter and anti-matter asymmetry, which is an important prediction of the Big Bang model. These significances have maintained CPV research at the forefront of both theoretical and experimental investigations since its initial proposal.

The discovery of CPV in hadrons can be traced back to 1964, when CPV was first observed in Kaon decays~\cite{Christenson:1964fg}. Since then, CPVs in mesons have gradually been discovered~\cite{BaBar:2001ags,Belle:2001zzw,CDF:2011ubb,CDF:2014pzb,LHCb:2019hro,Belle:2023str}. With the continuous accumulation of experimental data, it is natural for people to shift the focus of CPV research from mesons to baryons. However, the CPV of baryon was not discovered as expected until last year. In recent years, the signs of baryon CPV have emerged through LHCb experiments~\cite{LHCb:2024tnq,LHCb:2024yzj,LHCb:2024iis,LHCb:2025bfy}. Capitalizing on the LHC's high collision energies, the LHCb provides the ideal platform for observing b-baryon CPV. In 2025, LHCb first observed baryon CPV in $\Lambda_b$ four body decay processes~\cite{LHCb:2025ray}. This discovery directly fills the gap in the absence of CPV in baryons and presents new opportunities for the study of flavor physics.

In the past decade, the theoretical study of CPV, including the mechanism of generation and the detection method, has achieved considerable development~\cite{Gronau:2013mza,Hsiao:2014mua,He:2015fsa,Zhu:2016bra,He:2018joe,Roy:2019cky,Wang:2019dls,Roy:2020nyx,Wang:2022nbm,Dery:2022zkt,Wang:2022tcm,Bediaga:2022sxw,Shen:2023eln,Shen:2023nuw,Wang:2024oyi,Wang:2024qff,Sun:2024mmk,Jia:2024pyb,He:2024pxh,Cheng:2025kpp,Song:2025lmj}. Among various methods, the PQCD approach, as a perturbative method which is widely used in meson decays and capable of accurately predicting CPV, has gradually been applied to baryon decays and their CPV~\cite{Lu:2009cm,Rui:2022jff,Zhang:2022iun,Rui:2022sdc,Rui:2023fiz}. Recent work shows that the partial wave CPV in baryons has unusual offset mechanisms distinct from meson decays~\cite{Han:2024kgz}. Despite these advances, PQCD faces significant computational challenges and extended research cycles, particularly in baryon decays where the specific decay processes may has its specific mechanism. In this step, the PQCD cannot provide rapid guidance for baryon CPV experiments. Thus, complementary non-perturbative methods capable of preliminary predictions are essential. 

In CPV studies, the U-spin symmetry is a powerful tool~\cite{Gronau:2000zy,Gronau:2013mda,Bhattacharya:2015uua,Grossman:2018ptn,Dery:2021mll,Gavrilova:2022hbx,Schacht:2022kuj}. Since only the symmetry of two light down-type quarks is considered, the symmetry transformation will not affect the generation mechanism of CPV: the weak phase difference between tree level and penguin level. Last year, a new systemic U-spin symmetry method was proposed~\cite{Wang:2024rwf}. In this method, the amplitude can be clearly expressed by the U-spin matrix. This method systematically summarizes the U-spin analysis method and makes it possible to conduct more dynamic analysis on it by specifying the amplitude.
However, the Hamiltonian in ref.~\cite{Wang:2024rwf} is simply expressed by a doublet. Considering the penguin operator, the Hamiltonian can be decomposed as $2\otimes 2\otimes 2=4\oplus 2\oplus 2$, contradicting our previous analysis. To improve the theoretical foundation of this method, the strict decomposition will be given in this work.

Recently, we noticed some U-spin analysis based on the measured baryon CPV~\cite{Zhang:2025jnw,He:2025msg} align with our previous study~\cite{Wang:2024rwf}, using the CPV of $\Lambda_b$ decays, the U-spin can only give the CPV of $\Xi_b$ decays without any assumption. Fortunately, due to the complexity of four-body decays, the CPV of different decay structure like ${\cal A}_{CP}(\Lambda_b\to R(pK^-)R(\pi^+\pi^-))$ and ${\cal A}_{CP}(\Lambda_b\to R(p\pi^+\pi^-)K^-)$ are also measured.
Therefore, decay-structure analysis through U-spin symmetry may yield valuable insights.

The rest of this paper is organized as follows. In Sec.II the theoretical framework of U-spin including the decomposition of penguin operator is given. The systemic analysis of b-baryon four-body CPV under U-spin is presented in Sec.III. As we expect, the simple U-spin analysis cannot provide effective predictions. Therefore, the preliminary analysis involving dynamics will be given in Sec.IV. Based on the dynamic analysis we can give the effective prediction. Conclusions are given in the final section. 

%%%%%%%%%%%%%%%%%%%%%%%%
\section{theoretical framework}
%%%%%%%%%%%%%%%%%%%%%%%%
In U-spin symmetry, the b-baryon anti-triplet, light baryon octet and meson octet can be expressed by matrix~\cite{Wang:2024rwf} as
\begin{eqnarray}\label{ma}
B_{b}^1&=&(\Xi_b^-),\;B_{b}^2=(\Lambda_b^0,\Xi_b^0), \;B^1=\frac{\sqrt{3}}{2}\Sigma^0-\frac{1}{2}\Lambda^0,
M^2=(\pi^-,K^-),\;M^{\bar 2}=(\pi^+,K^+),\;B^{2}_2=(\Sigma^-,\Xi^-),\notag\\
M^3&=&\begin{pmatrix}
\frac{\sqrt{3}\eta_8}{2\sqrt{2}}-\frac{\pi^0}{2\sqrt{2}}&K^0 \\
\bar K^0&\frac{\pi^0}{2\sqrt{2}}-\frac{\sqrt{3}\eta_8}{2\sqrt{2}} \\
\end{pmatrix},\;B^{2}_1=(p,\Sigma^+),B^3=
\begin{pmatrix}
-\frac{\Sigma^0}{2\sqrt{2}}-\frac{\sqrt{3}\Lambda^0}{2\sqrt{2}}& n \\
\Xi^0&\frac{\Sigma^0}{2\sqrt{2}}+\frac{\sqrt{3}\Lambda^0}{2\sqrt{2}}\end{pmatrix}\end{eqnarray}
where the $\Sigma^-, \Xi^-$ can also be expressed by matrix $B^{ijk}_{S/A}$ as $B^{ijk}_{A}=\epsilon^{ij1}(B^2_2)^k$ and $B^{ijk}_{S}=\epsilon^{ki1}(B^2_2)^j+\epsilon^{kj1}(B^2_2)^i$ which is similar with the $SU(3)$ representation in Ref.~\cite{Wang:2024ztg}.  The baryon triplet can be expressed as $(B^3)^{ij}=\epsilon^{ik}(B^3)^j_k$ and $(B^1)^{ij}=B^1\epsilon^{ij}$. The details of constructing the
 matrices are given in Appendix \ref{appendixA}

Besides, the initial and final states in decays processes, the Hamiltonian can also be decomposed by the SU(2) group. To describe the bottom baryon charmless three-body decays, the low-energy effective weak Hamiltonian should also be represented by a U-spin matrix. The Hamiltonian that describes the $\Delta S = 1$ or $\Delta S = 0$ $B$ decays is~\cite{Buchalla:1995vs}
\begin{align}
{\cal H}_{\text{eff}} = \frac{G_F}{\sqrt{2}} \bigg[ V^*_{ub}V_{uq} \bigg( \sum^2_{i=1} C_i Q_i^{uq} + \sum^{10}_{i=3} C_i Q_i^q \bigg) + V^*_{cb}V_{cq} \bigg( \sum^2_{i=1} C_i Q_i^{cq} + \sum^{10}_{i=3} C_i Q_i^q \bigg) \bigg], \quad q = d, s,\label{HH}
\end{align}
where $C_i$ are the Wilson coefficients and $Q_i$ are the four-quark operators. Since we will focus on the U-spin symmetry of the Hamiltonian, the operators can be written by omitting their Dirac structure as
\begin{align}
Q_{1,2}^{qs} &\sim \bar b q \bar q s,\; Q_{1,2}^{qd} \sim \bar b q \bar q d, \; Q^s_{3-6} \sim \bar b s \sum \bar q^\prime q^\prime, Q^d_{3-6} \sim \bar b d \sum \bar q^\prime q^\prime, \; Q^s_{7-10} \sim \bar b s \sum e_{q^\prime} \bar q^\prime q^\prime, Q^d_{7-10} \sim \bar b d \sum e_{q^\prime} \bar q^\prime q^\prime,
\end{align}
where $q = u, d, s$.
The $Q_{1,2}$ represent the current-current operators, and $Q_{3-10}$ represent the penguin operators. Due to U-spin symmetry, the operators can be expressed as
\begin{align}
Q_{1,2}^{us} &\sim (\bar b u)(\bar u s), \quad Q_{1,2}^{cs} \sim (\bar b c)(\bar c s), \quad Q_{1,2}^{ud} \sim (\bar b u)(\bar u d), \quad Q_{1,2}^{cd} \sim (\bar b c)(\bar c d), \notag \\
Q^s_{3-10} &\sim \bar b s \sum \bar q^\prime q^\prime \sim (b \bar{s})(u \bar{u}) + (b \bar{s})(d \bar{d}) + (b \bar{s})(s \bar{s}),\quad Q^d_{3-10} \sim \bar b d \sum \bar q^\prime q^\prime \sim (b \bar{d})(u \bar{u}) + (b \bar{d})(d \bar{d}) + (b \bar{d})(s \bar{s}).
\end{align}
The tree-level operators $Q_{1,2}$ and the first terms of the penguin-level operators $Q_{3-10}$ can be regarded as forming a U-spin doublet as
$H^\prime_u = \{V^*_{ub}V_{ud}, V^*_{ub}V_{us}\}$ and $ H^\prime_c = \{V^*_{cb}V_{cd}, V^*_{cb}V_{cs}\}$.
By considering the flavor structure of quarks, the matrix elements of the Hamiltonian $H_k^{ij}$ for the latter two terms of the penguin operators can be defined as $\sum V_{\text{CKM}} Q^\prime_{3-10} = H_k^{ij} \times (\bar{b}q_i)(\bar{q_k}q_j)$,
where $\{i, j, k\} = \{1, 2\}$ with $\{q_1, q_2, q_3\} = \{d, s\}$. The Hamiltonian is decomposed as: $2 \otimes \bar{2} \otimes 2 = 4 \oplus 2 \oplus 2$. Following the decomposition of the SU(2) group, the decomposition can be expressed by through the following relations:
\begin{align}
H^{ij}_{k} &= \frac{1}{2} H(4)^{ij}_{k} - \frac{1}{3} H(2)^{ij}_{k} + \frac{2}{3} H(2^\prime)^{ij}_{k}, \notag \\
(H_4)^{ij}_{k} &= -\frac{1}{3} \big( H^{im}_m \delta^j_k + H^{jm}_m \delta^i_k + H^{mi}_m \delta^j_k + H^{mj}_m \delta^i_k \big) + H^{ij}_k + H^{ji}_k, \notag \\
(H_2)^{ij}_{k} &= H^{mi}_m \delta^j_k + H^{jm}_m \delta^i_k, \quad (H_2^\prime)^{ij}_{k} = H^{im}_m \delta^j_k + H^{mj}_m \delta^i_k.
\end{align}
Based on the aforementioned decomposition relationship, we can derive the Hamiltonian matrix in irreducible representation amplitudes (IRA) method corresponding to each operator. The Hamiltonian matrix are $H^{21}_{1}=H^{22}_{2}= V^*_{us}V_{ub} +  V^*_{cs}V_{cb}$ and $ H^{12}_{2}=H^{11}_{1}= V^*_{ud}V_{ub} +  V^*_{cd}V_{cb}$ . Therefore, the IRA Hamiltonian can be derived as  
\begin{align}
2(H_2)^{11}_{1} &= 3(H_2)^{21}_{2} = 6(H_2)^{12}_{2} = 2 V^*_{ud}V_{ub} + 2 V^*_{cd}V_{cb}, \quad 2(H_2)^{22}_{2} = 3(H_2)^{12}_{1} = 6(H_2)^{21}_{1} = 2 V^*_{us}V_{ub} + 2 V^*_{cs}V_{cb}, \notag \\
(H_2)^{1} &= \frac{5}{3} \big( V^*_{ud}V_{ub} + V^*_{cd}V_{cb} \big),\quad  (H_2)^{2} = \frac{5}{3} \big( V^*_{us}V_{ub} + V^*_{cs}V_{cb} \big),\qquad (H_4) = 0, 
\end{align}
where $(H_2)^i = (H_2)^{ji}_k \delta^k_j$. Through group decomposition, we find that $H_2$ remains a U-spin doublet and can be identified as: $(H_2)_u =  \frac{5}{3}\{V^*_{ub}V_{ud}, V^*_{ub}V_{us}\}$ and $ (H_2)_c =  \frac{5}{3}\{V^*_{cb}V_{cd}, V^*_{cb}V_{cs}\}$. Then the Hamiltonian including the tree and penguin level can be written in U-spin as ${\cal H}_{\text{eff}} =\{H^\prime_u + (H_2)_u\} + \{H^\prime_c + (H_2)_c\} = \frac{8}{3}(H_u + H_c)$, and $H_{u,c}$ form the doublet, respectively, as
\begin{align}
H_u &= (V^*_{ub}V_{ud}, V^*_{ub}V_{us}), \quad H_c = (V^*_{cb}V_{cd}, V^*_{cb}V_{cs}).
\end{align}
Therefore, although the penguin operator will introduce different SU(2) decompositions as $2 \otimes \bar{2} \otimes 2 = 4 \oplus 2 \oplus 2$, considering the specific Hamiltonian expression, the $4$ will not contribute to our analysis and the doublet from the penguin operator can be absorbed into the tree-level doublet  Hamiltonian. Thus only the doublet Hamiltonian can be involved in our analysis and our previous study~\cite{Wang:2024rwf} is correct. 

%%%%%%%%%%%%%%%%%%%%%%%%
\section{ U-spin analysis of b-baryon four body decay CPV}
%%%%%%%%%%%%%%%%%%%%%%%%
According to the U-spin symmetry, charmless four-body decays of a bottom baryon can then be divided into eleven types as
\begin{eqnarray}
    &B_b^2 \to B^{3/1} M^3 M^3 M^3 ,B_b^2 \to B^{3/1} M^3 M^2 M^{\bar 2}, B_b^2 \to B^2_1 M^2 M^2 M^{\bar{2}}, B_b^2 \to B^2_1 M^2 M^3 M^3 , \notag \\
    &B_b^2 \to B^2_2 M^3 M^3 M^{\bar{2}}, B_b^2 \to B^2_2 M^{\bar{2}}  M^{\bar{2}} M^2, B_b^1 \to B^{3/1} M^3 M^3 M^2,B_b^1 \to B^{3/1} M^2 M^2 M^{\bar 2}, \notag\\ 
    &B_b^1 \to B^2_2 M^3 M^3 M^3,B_b^1 \to B^2_2 M^2 M^{\bar 2} M^3, B_b^1 \to B^2_1 M^2 M^2 M^3 .
\end{eqnarray}
This also reflects the problem of U-spin analysis comparing to SU(3) flavor analysis. The reduced symmetry will bring too many amplitude types and the predictive power will be decreased. But in CPV studies, the SU(3) flavor cannot replace the U-spin analysis since the introduction of up-type quarks will break the symmetry in the CKM matrix. This symmetry enables the derivation of CPV relations under U-spin analysis. The detailed discussion can be found in ref.~\cite{Wang:2024rwf}. 
The corresponding amplitudes of bottom baryon doublet decays can be built as
\begin{align}
{\cal A}_{3333}^{2} &= \sum_{q=u,c} \bigg( a_{1q} (B_{b}^{2})^{i} (H_{q})^{k} (B^{3})_{ij} (M^{3})^{j}_{k} (M^{3})^{m}_{l} (M^{3})^{l}_{m}
+ a_{2q} (B_{b}^{2})^{k} (H_{q})^{i} (B^{3})_{ij} (M^{3})^{j}_{k} (M^{3})^{m}_{l} (M^{3})^{l}_{m} \bigg), \notag \\
%%%%%%%%%%%
{\cal A}_{1333}^{2} &= \sum_{q=u,c} ~ a^1_{1q} (B_{b}^{2})^{i} (H_{q})^{k} (B^{1})_{ij} (M^{3})^{j}_{k} (M^{3})^{m}_{l} (M^{3})^{l}_{m}, \notag \\
%%%%%%%%%%%
{\cal A}_{332 \bar2}^{2} &= \sum_{q=u,c} \bigg( b_{1q} (B_{b}^{2})^{i} (H_{q})^{j} (B^{3})_{ij} (M^{3})^{l}_{k} (M^{2})_{l} (M^{\bar{2}})^k
+ b_{2q} (B_{b}^{2})^{i} (H_{q})^{k} (B^{3})_{ij} (M^{3})^{j}_{k} (M^{2})_{l} (M^{\bar{2}})^l \notag \\
&\quad \quad \quad  + b_{3q} (B_{b}^{2})^{i} (H_{q})^{k} (B^{3})_{ij} (M^{3})^{l}_{k} (M^{2})_{l} (M^{\bar{2}})^j 
+ b_{4q} (B_{b}^{2})^{i} (H_{q})^{l} (B^{3})_{ij} (M^{3})^{j}_{k} (M^{2})_{l} (M^{\bar{2}})^k \notag \\
&\quad \quad \quad  + b_{5q} (B_{b}^{2})^{k} (H_{q})^{i} (B^{3})_{ij} (M^{3})^{j}_{k} (M^{2})_{l} (M^{\bar{2}})^l 
+ b_{6q} (B_{b}^{2})^{k} (H_{q})^{i} (B^{3})_{ij} (M^{3})^{l}_{k} (M^{2})_{l} (M^{\bar{2}})^j \notag \\
&\quad \quad \quad  + b_{7q} (B_{b}^{2})^{k} (H_{q})^{l} (B^{3})_{ij} (M^{3})^{i}_{k} (M^{2})_{l} (M^{\bar{2}})^j 
+ b_{8q} (B_{b}^{2})^{l} (H_{q})^{i} (B^{3})_{ij} (M^{3})^{j}_{k} (M^{2})_{l} (M^{\bar{2}})^k \notag \\
&\quad \quad \quad  + b_{9q} (B_{b}^{2})^{l} (H_{q})^{k} (B^{3})_{ij} (M^{3})^{i}_{k} (M^{2})_{l} (M^{\bar{2}})^j \bigg) ,\notag \\
%%%%%%%%%%%%%%%
{\cal A}_{132 \bar2}^{2} &= \sum_{q=u,c} \bigg( b_{1q}^1 (B_{b}^{2})^{i} (H_{q})^{j} (B^{3})_{ij} (M^{3})^{l}_{k} (M^{2})_{l} (M^{\bar{2}})^k
+ b_{2q}^1 (B_{b}^{2})^{i} (H_{q})^{k} (B^{1})_{ij} (M^{3})^{j}_{k} (M^{2})_{l} (M^{\bar{2}})^l \notag \\
&\quad \quad \quad  + b_{3q}^1 (B_{b}^{2})^{i} (H_{q})^{k} (B^{1})_{ij} (M^{3})^{l}_{k} (M^{2})_{l} (M^{\bar{2}})^j 
+ b_{4q}^1 (B_{b}^{2})^{i} (H_{q})^{l} (B^{1})_{ij} (M^{3})^{j}_{k} (M^{2})_{l} (M^{\bar{2}})^k \notag \\
&\quad \quad \quad  + b_{5q}^1 (B_{b}^{2})^{k} (H_{q})^{i} (B^{1})_{ij} (M^{3})^{l}_{k} (M^{2})_{l} (M^{\bar{2}})^j 
+ b_{6q}^1 (B_{b}^{2})^{l} (H_{q})^{i} (B^{1})_{ij} (M^{3})^{j}_{k} (M^{2})_{l} (M^{\bar{2}})^k \bigg) ,\notag \\
%%%%%%%%%%%%%
{\cal A}_{222 \bar2}^{2} &= \sum_{q=u,c} \bigg( c_{1q} (B_{b}^2)^i (H_q)^j (B^2_1)_{i} (M^{2})_{j} (M^{2})_{k}(M^{\bar{2}})^k
+ c_{2q} (B_{b}^2)^j (H_q)^i (B^2_1)_{i} (M^{2})_{j} (M^{2})_{k}(M^{\bar{2}})^k \notag \\
&\quad  \quad  \quad + c_{3q} (B_{b}^2)^j (H_q)^k (B^2_1)_{i} (M^{2})_{j} (M^{2})_{k}(M^{\bar{2}})^i \bigg), \notag \\
%%%%%%%%%%
{\cal A}_{222 3}^{2} &= \sum_{q=u,c} \bigg( d_{1q} (B_{b}^2)^i (H_q)^j (B^2_1)_{i} (M^{2})_{j} (M^{3})^{l}_{k} (M^{3})^{k}_{l}
+ d_{2q} (B_{b}^2)^j (H_q)^i (B^2_1)_{i} (M^{2})_{j} (M^{3})^{l}_{k} (M^{3})^{k}_{l} \notag \\
&\quad  \quad  \quad + d_{3q} (B_{b}^2)^k (H_q)^l (B^2_1)_{i} (M^{2})_{j} (M^{3})^{i}_{k} (M^{3})^{j}_{l}
+ d_{4q} (B_{b}^2)^k (H_q)^l (B^2_1)_{i} (M^{2})_{j} (M^{3})^{j}_{k} (M^{3})^{i}_{l} \bigg) ,\notag \\
%%%%%%%%%%%%%%%%%%
{\cal A}_{2 \bar2 3 3 }^{2} &= \sum_{q=u,c} \bigg( e_{1q} (B_{b}^2)^i (H_q)^j (B_S)_{ijk} (M^{\bar{2}})^k (M^{3})^{m}_{l} (M^{3})^{l}_{m}
+ e_{3q} (B_{b}^2)^i (H_q)^l (B_S)_{ijk} (M^{\bar{2}})^m (M^{3})^{j}_{l} (M^{3})^{k}_{m} \notag \\
&\quad  \quad  \quad + e_{4q} (B_{b}^2)^i (H_q)^l (B_S)_{ijk} (M^{\bar{2}})^m (M^{3})^{k}_{l} (M^{3})^{j}_{m}
+ e_{5q} (B_{b}^2)^k (H_q)^i (B_S)_{ijk} (M^{\bar{2}})^j (M^{3})^{m}_{l} (M^{3})^{l}_{m} \notag \\
&\quad  \quad  \quad + e_{6q} (B_{b}^2)^k (H_q)^l (B_S)_{ijk} (M^{\bar{2}})^m (M^{3})^{i}_{l} (M^{3})^{j}_{m} 
+ e_{7q} (B_{b}^2)^l (H_q)^i (B_S)_{ijk} (M^{\bar{2}})^m (M^{3})^{j}_{l} (M^{3})^{k}_{m} \notag \\
&\quad  \quad  \quad + e_{8q} (B_{b}^2)^l (H_q)^i (B_S)_{ijk} (M^{\bar{2}})^m (M^{3})^{k}_{l} (M^{3})^{j}_{m}
+ e_{9q} (B_{b}^2)^l (H_q)^k (B_S)_{ijk} (M^{\bar{2}})^m (M^{3})^{i}_{l} (M^{3})^{j}_{m} \notag \\
&\quad  \quad  \quad + e_{10q} (B_{b}^2)^l (H_q)^m (B_S)_{ijk} (M^{\bar{2}})^i (M^{3})^{j}_{l} (M^{3})^{k}_{m}
+ e_{11q} (B_{b}^2)^l (H_q)^m (B_S)_{ijk} (M^{\bar{2}})^i (M^{3})^{m}_{k} (M^{3})^{j}_{m} \notag \\
&\quad  \quad  \quad + e_{12q} (B_{b}^2)^l (H_q)^m (B_S)_{ijk} (M^{\bar{2}})^k (M^{3})^{i}_{k} (M^{3})^{j}_{m}
+ e_{13q} (B_{b}^2)^i (H_q)^j (B_S)_{ijk} (M^{\bar{2}})^k (M^{3})^{m}_{l} (M^{3})^{l}_{m} \notag \\
&\quad  \quad  \quad + e_{14q} (B_{b}^2)^i (H_q)^k (B_A)_{ijk} (M^{\bar{2}})^j (M^{3})^{m}_{l} (M^{3})^{l}_{m}
+ e_{15q} (B_{b}^2)^i (H_q)^l (B_A)_{ijk} (M^{\bar{2}})^m (M^{3})^{j}_{l} (M^{3})^{k}_{m} \notag \\
&\quad  \quad  \quad + e_{16q} (B_{b}^2)^i (H_q)^l (B_A)_{ijk} (M^{\bar{2}})^m (M^{3})^{k}_{l} (M^{3})^{j}_{m}
+ e_{17q} (B_{b}^2)^k (H_q)^i (B_A)_{ijk} (M^{\bar{2}})^j (M^{3})^{m}_{l} (M^{3})^{l}_{m} \notag \\
&\quad  \quad  \quad + e_{18q} (B_{b}^2)^l (H_q)^i (B_A)_{ijk} (M^{\bar{2}})^m (M^{3})^{k}_{l} (M^{3})^{j}_{m} \bigg), \notag \\
%%%%%%%%%%%%%%%%
{\cal A}_{2\bar2 \bar2 2}^{2} &= \sum_{q=u,c} \bigg( f_{1q} (B_{b}^2)^i (H_q)^j (B_S)_{ijk} (M^{\bar{2}})^k (M^{\bar{2}})^l (M^{2})_{l}
+ f_{2q} (B_{b}^2)^i (H_q)^k (B_S)_{ijk} (M^{\bar{2}})^j (M^{\bar{2}})^l (M^{2})_{l} \notag \\
&\quad  \quad  \quad + f_{3q} (B_{b}^2)^i (H_q)^l (B_S)_{ijk} (M^{\bar{2}})^j (M^{\bar{2}})^k (M^{2})_{l}
+ f_{4q} (B_{b}^2)^k (H_q)^i (B_S)_{ijk} (M^{\bar{2}})^j (M^{\bar{2}})^l (M^{2})_{l} \notag \\
&\quad  \quad  \quad + f_{5q} (B_{b}^2)^l (H_q)^i (B_S)_{ijk} (M^{\bar{2}})^j (M^{\bar{2}})^k (M^{2})_{l}
+ f_{6q} (B_{b}^2)^i (H_q)^j (B_A)_{ijk} (M^{\bar{2}})^k (M^{\bar{2}})^l (M^{2})_{l} \notag \\
&\quad  \quad  \quad + f_{7q} (B_{b}^2)^i (H_q)^k (B_A)_{ijk} (M^{\bar{2}})^j (M^{\bar{2}})^l (M^{2})_{l}
+ f_{8q} (B_{b}^2)^i (H_q)^l (B_A)_{ijk} (M^{\bar{2}})^j (M^{\bar{2}})^k (M^{2})_{l} \notag \\
&\quad  \quad  \quad + f_{9q} (B_{b}^2)^k (H_q)^i (B_A)_{ijk} (M^{\bar{2}})^j (M^{\bar{2}})^l (M^{2})_{l}
+ f_{10q} (B_{b}^2)^l (H_q)^i (B_A)_{ijk} (M^{\bar{2}})^j (M^{\bar{2}})^k (M^{2})_{l} \bigg).\label{triplet}
\end{align}

As for the singlet which including $\Lambda^0$ and $\Sigma^0$ ,the corresponding amplitudes decays can be built as
\begin{align}
\mathcal{A}_{3332}^1 &= \sum_{q=u,c} \bigg( A_{1q} (H_q)^{i} (B^3)^{j}_{i} (M^3)^{k}_{j} (M^3)^{l}_{k} (M^2)_{l}
+ A_{2q} (H_q)^{j} (B^3)^{k}_{i} (M^3)^{i}_{j} (M^3)^{l}_{k} (M^2)_{l} \notag \\
&\quad \quad \quad + A_{3q} (H_q)^{k} (B^3)^{j}_{i} (M^3)^{i}_{j} (M^3)^{l}_{k} (M^2)_{l} \bigg), \notag \\
%%%%%%%%%%%%%%%
\mathcal{A}_{1332}^1 &= \sum_{q=u,c} A^1_{1q} (H_q)^{i} (B^1)^{j}_{i} (M^3)^{k}_{j} (M^3)^{l}_{k} (M^2)_{l} , \notag \\
 %%%%%%%%%%%%%%%
\mathcal{A}_{322\bar{2}}^1 &= \sum_{q=u,c} \bigg( B_{1q} (H_q)^{i} (B^3)^{j}_{i} (M^2)_{j} (M^2)_{k} (M^{\bar{2}})^{k}
+ B_{2q} (H_q)^{j} (B^3)^{k}_{i} (M^2)_{j} (M^2)_{k} (M^{\bar{2}})^{i} \bigg), \notag \\
%%%%%%%%%%%%%%%
\mathcal{A}_{122\bar{2}}^1 &= \sum_{q=u,c}  B^1_{1q} (H_q)^{i} (B^1)^{j}_{i} (M^2)_{j} (M^2)_{k} (M^{\bar{2}})^{k} , \notag \\
%%%%%%%%%%%%%%%
\mathcal{A}_{2322}^1 &= \sum_{q=u,c} \bigg( C_{1q} (B^1_b)^{jl} (H_q)^{i} (B^2_1)_{i} (M^3)^{k}_{j} (M^2)_{k} (M^2)_{l}
 + C_{2q} (B^1_b)^{jl} (H_q)^{k} (B^2_1)_{i} (M^3)^{i}_{j} (M^2)_{k} (M^2)_{l} \notag \\
&\quad \quad \quad + C_{3q} (B^1_b)^{il} (H_q)^{j} (B^2_1)_{i} (M^3)^{k}_{j} (M^2)_{k} (M^2)_{l} \bigg), \notag \\
%%%%%%%%%%%%%%%
\mathcal{A}_{232\bar{2}}^1 &= \sum_{q=u,c} \bigg( D_{1q} (H_q)^{i} (B^2_2)_{i} (M^3)^{k}_{j} (M^2)_{k} (M^{\bar{2}})^{j}
+ D_{2q} (H_q)^{j} (B^2_2)_{i} (M^3)^{i}_{j} (M^2)_{k} (M^{\bar{2}})^{k} \notag \\
&\quad \quad \quad + D_{3q} (H_q)^{k} (B^2_2)_{i} (M^3)^{i}_{j} (M^2)_{k} (M^{\bar{2}})^{j}
+ D_{4q} (H_q)^{j} (B^2_2)_{i} (M^3)^{k}_{j} (M^2)_{k} (M^{\bar{2}})^{i} \bigg), \notag \\
%%%%%%%%%%%%%%%
\mathcal{A}_{2333}^1 &= \sum_{q=u,c} \bigg( E_{1q} (H_q)^{j} (B^2_2)_{i} (M^3)^{i}_{j} (M^3)^{l}_{k} (M^3)^{k}_{l} \bigg).\label{singlet}
\end{align}

We noticed that when we consider the angular momentum between two identical meson in the final state, the amplitude must be symmetry(anti-symmetry) under the exchange of the SU(3) quantum number of  two identical meson with even(odd)  angular momentum~\cite{Savage:1989qr}. Since the amplitude in Eq.\eqref{triplet} and Eq.\eqref{singlet} are symmetry under the exchange of the SU(3) quantum number of  two identical meson, we only consider the S wave of the angular momentum between two identical meson and omit the higher angular momentum.  

The corresponding amplitudes of specific b-baryon decays can be derived by expending these amplitudes in Eq.\eqref{triplet} and Eq.\eqref{singlet}. The amplitude of each b-baryon four body decays are given in Table.\ref{table1}, Table.\ref{table2} and Table.\ref{table3}.

\begin{table*}[htbp!]
    \caption{The U-spin amplitudes of charmless bottomed baryon four body decays. Here $\lambda_s^u=V_{ub}^*V_{us}$, $\lambda_d^u=V_{ub}^*V_{ud}$, $\lambda_s^c=V_{cb}^*V_{cs}$ and $\lambda_d^c=V_{cb}^*V_{cd}$. The $\sum_q$ means sum the index $q$ with $q=u,c$.}\label{table1}
    \renewcommand{\arraystretch}{1.25}
    \begin{tabular}{cccc}\hline\hline
    channel & amplitude & channel & amplitude \\\hline 
$\Xi_b^0\to nK^0  \bar{K}^{0} \bar{K}^0$ & $ \sum_q \lambda_d^q  a_2$ & $\Lambda_b^0\to \Xi^0K^0  \bar{K}^0 K^0$ & $ \sum_q \lambda_s^q  a_2$ \tabularnewline\hline
$\Xi_b^0\to \Sigma^0K^0  \bar{K}^0 \bar{K}^0$ & $ \sum_q \lambda_s^q (\frac{1}{2 \sqrt{2}}( a_1+a_2)+\frac{\sqrt{3}}{2}a_1^1)$ & $\Lambda_b^0\to \Sigma^0K^0  \bar{K}^0 K^0$ & $ \sum_q \lambda_d^q (\frac{1}{2 \sqrt{2}}( a_1+a_2)-\frac{\sqrt{3}}{2}a_1^1)$ \tabularnewline\hline
$\Xi_b^0\to \Lambda^0K^0  \bar{K}^0 \bar{K}^0$ & $ \sum_q \lambda_s^q  (\frac{\sqrt{3}}{2 \sqrt{2}} \left( a_1+a_2\right)-\frac{1}{2}a_1^1)$ & $\Lambda_b^0\to \Lambda^0K^0  \bar{K}^0 K^0$ & $ \sum_q \lambda_d^q (\frac{\sqrt{3}}{2\sqrt{2}} \left( a_1+a_2\right)+\frac{1}{2}a_1^1)$ \tabularnewline\hline
$\Xi_b^0\to \Xi^0 K^0  \bar{K}^0 K^0$ & $ \sum_q\lambda_d^q a_1$ & $\Lambda_b^0\to n K^0  \bar{K}^0 \bar{K}^0$ & $ \sum_q\lambda_s^q a_1$ \tabularnewline\hline
\hline
$\Lambda_b^0\to \Sigma^0 K^0  \pi^- \pi^+$ & $\begin{aligned} & \sum_q\lambda_d^q(\frac{1}{2 \sqrt{2}}(b_2+b_4+b_5+b_7+b_8+b_9) \\ & +\frac{\sqrt{3}}{2}(b_2^1+b_4^1+b_6^1)) \end{aligned}$ & $\Xi_b^0\to \Sigma^0 \bar{K}^0  K^- K^+$ & $\begin{aligned} & \sum_q\lambda_s^q(\frac{1}{2 \sqrt{2}}(b_2+b_4+b_5+b_7+b_8+b_9) \\ & +\frac{\sqrt{3}}{2}(b_2^1+b_4^1+b_6^1)) \end{aligned}$ \tabularnewline\hline
$\Lambda_b^0\to \Sigma^0 K^0  K^- K^+$ & $\begin{aligned} & \sum_q\lambda_d^q(\frac{1}{2 \sqrt{2}}(b_2+b_3+b_5+b_6) \\ & +\frac{\sqrt{3}}{2}(b_2^1+b_3^1+b_5^1)) \end{aligned}$ & $\Xi_b^0\to \Sigma^0 \bar{K}^0  \pi^- \pi^+$ & $\begin{aligned} & \sum_q\lambda_s^q(\frac{1}{2 \sqrt{2}}(b_2+b_3+b_5+b_6) \\ & +\frac{\sqrt{3}}{2}(b_2^1+b_3^1+b_5^1)) \end{aligned}$ \tabularnewline\hline
$\Xi_b^0\to \Sigma^0 K^0  K^- \pi^+$ & $\begin{aligned} & \sum_q\lambda_d^q(\frac{1}{2 \sqrt{2}}(b_1+b_3+b_8+b_9) \\ & +\frac{\sqrt{3}}{2}(b_1^1+b_3^1-b_6^1)) \end{aligned}$ & $\Lambda_b^0\to \Sigma^0 \bar{K}^0  \pi^- K^+$ & $\begin{aligned} & \sum_q\lambda_s^q(\frac{1}{2 \sqrt{2}}(b_1+b_3+b_8+b_9) \\ & +\frac{\sqrt{3}}{2}(b_1^1+b_3^1-b_6^1)) \end{aligned}$ \tabularnewline\hline
$\Xi_b^0\to \Sigma^0 \bar{K}^0  \pi^- K^+$ & $\begin{aligned} & \sum_q\lambda_d^q(\frac{1}{2 \sqrt{2}}(b_1+b_4+b_6+b_7) \\ & +\frac{\sqrt{3}}{2}(b_1^1+b_4^1-b_5^1)) \end{aligned}$ & $\Lambda_b^0\to \Sigma^0 K^0  K^- \pi^+$ & $\begin{aligned} & \sum_q\lambda_s^q(\frac{1}{2 \sqrt{2}}(b_1+b_4+b_6+b_7) \\ & +\frac{\sqrt{3}}{2}(b_1^1+b_4^1-b_5^1)) \end{aligned}$ \tabularnewline\hline
$\Lambda_b^0\to \Lambda^0 K^0  \pi^- \pi^+$ & $\begin{aligned} & \sum_q\lambda_d^q(\frac{\sqrt{3}}{2\sqrt{2}}(b_2+b_4+b_5+b_7+b_8+b_9) \\ & -\frac{1}{2}(b_2^1+b_4^1+b_6^1)) \end{aligned}$ & $\Xi_b^0\to \Lambda^0 \bar{K}^0  K^- K^+$ & $\begin{aligned} & \sum_q\lambda_s^q(\frac{\sqrt{3}}{2\sqrt{2}}(b_2+b_4+b_5+b_7+b_8+b_9) \\ & +\frac{1}{2}(b_2^1+b_4^1+b_6^1)) \end{aligned}$ \tabularnewline\hline
$\Lambda_b^0\to \Lambda^0 K^0  K^- K^+$ & $\begin{aligned} & \sum_q\lambda_d^q(\frac{\sqrt{3}}{2\sqrt{2}}(b_2+b_3+b_5+b_6) \\ & -\frac{1}{2}(b_2^1+b_3^1+b_5^1)) \end{aligned}$ & $\Xi_b^0\to \Lambda^0 \bar{K}^0  \pi^- \pi^+$ & $\begin{aligned} & \sum_q\lambda_s^q(\frac{\sqrt{3}}{2\sqrt{2}}(b_2+b_3+b_5+b_6) \\ & +\frac{1}{2}(b_2^1+b_3^1+b_5^1)) \end{aligned}$ \tabularnewline\hline
$\Xi_b^0\to \Lambda^0 K^0  K^- \pi^+$ & $\begin{aligned} & \sum_q\lambda_d^q(\frac{\sqrt{3}}{2\sqrt{2}}(b_1+b_3+b_8+b_9) \\ & +\frac{1}{2}(b_1^1+b_3^1-b_6^1)) \end{aligned}$ & $\Lambda_b^0\to \Lambda^0 \bar{K}^0  \pi^- K^+$ & $\begin{aligned} & \sum_q\lambda_s^q(\frac{\sqrt{3}}{2\sqrt{2}}(b_1+b_3+b_8+b_9) \\ & +\frac{1}{2}(-b_1^1-b_3^1+b_6^1)) \end{aligned}$ \tabularnewline\hline
$\Xi_b^0\to \Lambda^0 \bar{K}^0  \pi^- K^+$ & $\begin{aligned} & \sum_q\lambda_d^q(\frac{\sqrt{3}}{2\sqrt{2}}(b_1+b_4+b_6+b_7) \\ & +\frac{1}{2}(b_1^1+b_4^1-b_5^1)) \end{aligned}$ & $\Lambda_b^0\to \Lambda^0 K^0  K^- \pi^+$ & $\begin{aligned} & \sum_q\lambda_s^q(\frac{\sqrt{3}}{2\sqrt{2}}(b_1+b_4+b_6+b_7) \\ & +\frac{1}{2}(-b_1^1-b_4^1+b_5^1)) \end{aligned}$ \tabularnewline\hline
$\Lambda_b^0\to \Xi^0 K^0  \pi^- K^+$ & $ \sum_q\lambda_d^q(b_7+b_9)$ & $\Xi_b^0\to n \bar{K}^0  K^- \pi^+$ & $ \sum_q\lambda_s^q(b_7+b_9)$ \tabularnewline\hline
$\Xi_b^0\to \Xi^0 K^0  \pi^- \pi^+$ & $ \sum_q\lambda_d^q(b_2+b_4)$ & $\Lambda_b^0\to n \bar{K}^0  K^- K^+$ & $ \sum_q\lambda_s^q(b_2+b_4)$ \tabularnewline\hline
$\Xi_b^0\to \Xi^0 K^0  K^- K^+$ & $ \sum_q\lambda_d^q(b_2+b_3+b_9)$ & $\Lambda_b^0\to n \bar{K}^0  \pi^- \pi^+$ & $ \sum_q\lambda_s^q(b_2+b_3+b_9)$ \tabularnewline\hline
$\Lambda_b^0\to n K^0  K^- \pi^+$ & $ \sum_q\lambda_d^q(b_1+b_3+b_6)$ & $\Xi_b^0\to \Xi^0 \bar{K}^0  \pi^- K^+$ & $ \sum_q\lambda_s^q(b_1+b_3+b_6)$ \tabularnewline\hline
$\Lambda_b^0\to n \bar{K}^0  \pi^- K^+$ & $ \sum_q\lambda_d^q(b_1+b_4+b_8)$ & $\Xi_b^0\to \Xi^0 K^0  K^- \pi^+$ & $ \sum_q\lambda_s^q(b_1+b_4+b_8)$ \tabularnewline\hline
$\Xi_b^0\to n \bar{K}^0  \pi^- \pi^+$ & $ \sum_q\lambda_d^q(b_5+b_6+b_7)$ & $\Lambda_b^0\to \Xi^0 K^0  K^- K^+$ & $ \sum_q\lambda_s^q(b_5+b_6+b_7)$ \tabularnewline\hline
$\Xi_b^0\to n \bar{K}^0  K^- K^+$ & $ \sum_q\lambda_d^q(b_5+b_8)$ & $\Lambda_b^0\to \Xi^0 K^0  \pi^- \pi^+$ & $ \sum_q\lambda_s^q(b_5+b_8)$ \tabularnewline\hline
    \hline
    \end{tabular}
    \end{table*}
%%%%%%%%%%%%%%%%%%%%%%%%%%%%%%%%%%%%%%%%%%%%%
%%%%%%%%%%%%%%%%%%%%%%%%%%%%%%%%%%%%%%%%%%%%%
\begin{table*}[htbp!]
    \caption{The U-spin amplitudes of charmless bottomed baryon four body decays. Here $\lambda_s^u=V_{ub}^*V_{us}$, $\lambda_d^u=V_{ub}^*V_{ud}$, $\lambda_s^c=V_{cb}^*V_{cs}$ and $\lambda_d^c=V_{cb}^*V_{cd}$. The $\sum_q$ means sum the index $q$ with $q=u,c$.}\label{table2}
    \renewcommand{\arraystretch}{1.35}
    \begin{tabular}{cccc}\hline\hline
    channel & amplitude & channel & amplitude \\\hline
    $\Lambda_b^0\to p \pi^-  \pi^- \pi^+$ & $ \sum_q\lambda_d^q(c_1+c_2+c_3)$ & $\Xi_b^0\to \Sigma^+ K^-  K^- K^+$ & $ \sum_q\lambda_s^q(c_1+c_2+c_3)$ \tabularnewline\hline
    $\Lambda_b^0\to p K^-  \pi^- K^+$ & $ \sum_q\lambda_d^q(c_1+c_2)$ & $\Xi_b^0\to \Sigma^+ K^-  \pi^- \pi^+$ & $ \sum_q\lambda_s^q(c_1+c_2)$ \tabularnewline\hline
    $\Xi_b^0\to p K^-  \pi^- \pi^+$ & $ \sum_q\lambda_d^q(c_2+c_3)$ & $\Lambda_b^0\to \Sigma^+ K^-  \pi^- K^+$ & $ \sum_q\lambda_s^q(c_2+c_3)$ \tabularnewline\hline
    $\Xi_b^0\to p K^-  K^- K^+$ & $ \sum_q\lambda_d^q c_2$ & $\Lambda_b^0\to \Sigma^+ \pi^-  \pi^- \pi^+$ & $ \sum_q\lambda_s^q c_2$ \tabularnewline\hline
    $\Lambda_b^0\to \Sigma^+ \pi^-  \pi^- K^+$ & $ \sum_q\lambda_d^q c_3$ & $\Xi_b^0\to p K^-  K^- \pi^+$ & $ \sum_q\lambda_s^q c_3$ \tabularnewline\hline
    $\Xi_b^0\to \Sigma^+ \pi^-  \pi^- \pi^+$ & $ \sum_q\lambda_d^q c_1$ & $\Lambda_b^0\to p K^-  K^- K^+$ & $ \sum_q\lambda_s^q c_1$ \tabularnewline\hline
    $\Xi_b^0\to \Sigma^+ \pi^-  K^- K^+$ & $ \sum_q\lambda_d^q(c_1+c_3)$ & $\Lambda_b^0\to p \pi^-  K^- \pi^+$ & $ \sum_q\lambda_s^q(c_1+c_3)$    \tabularnewline\hline
\hline
    $\Lambda_b^0\to p K^0 \bar{K}^0 \pi^-$ & $\sum_q\lambda_d^q( d_1+d_2)$ & $\Xi_b^0\to \Sigma^+ \bar{K}^0 K^0 K^-$ & $\sum_q\lambda_s^q(d_1+d_2)$ \tabularnewline\hline
    $\Xi_b^0\to p \bar{K}^0 K^0 K^-$ & $\sum_q\lambda_d^q(d_2+d_3)$ & $\Lambda_b^0\to \Sigma^+ K^0 \bar{K}^0 \pi^-$ & $\sum_q\lambda_s^q( d_2+d_3)$ \tabularnewline\hline
    $\Lambda_b^0\to \Sigma^+ K^0 K^0 K^-$ & $\sum_q\lambda_d^q(d_3+d_4)$ &$\Xi_b^0\to p \bar{K}^0 \bar{K}^0 \pi^-$ & $\sum_q\lambda_s^q(d_3+d_4)$ \tabularnewline\hline
    $\Xi_b^0\to \Sigma^+ K^0 \bar{K}^0 \pi^-$ & $\sum_q\lambda_d^q( d_1+d_4)$ & $\Lambda_b^0\to p \bar{K}^0 K^0 K^-$ & $\sum_q\lambda_s^q( d_1+d_4)$ \tabularnewline\hline
    \hline
    $\Lambda_b^0\to \Sigma^- K^+  K^0 \bar{K}^0$ & 
    $\begin{aligned} 
    & \sum_q\lambda_d^q(-4 e_1+2 e_2+e_3-2 e_4+2 e_5 \\ 
    & +e_6+e_7-2 e_8+e_9+2 e_{14} \\ 
    & +e_{15}+2 e_{17}) 
    \end{aligned}$ &
    $\Xi_b^0\to \Xi^- \pi^+  \bar{K}^0 K^0$ & 
    $\begin{aligned} 
    & \sum_q\lambda_s^q(4 e_1-2 e_2-e_3+2 e_4-2 e_5 \\ 
    & -e_6-e_7+2 e_8-e_9-2 e_{14} \\ 
    & -e_{15}-2 e_{17}) 
    \end{aligned}$ \tabularnewline\hline
    
    $\Xi_b^0\to \Sigma^- \pi^+  \bar{K}^0 K^0$ & 
    $\begin{aligned} 
    & \sum_q\lambda_d^q(2 e_1+2 e_2-4 e_5 \\ 
    & -2 e_7+e_8+e_9-2 e_{10} \\ 
    & +e_{11}+e_{12}-2 e_{13}-2 e_{14}+e_{18}) 
    \end{aligned}$ &
    $\Lambda_b^0\to \Xi^- K^+  K^0 \bar{K}^0$ & 
    $\begin{aligned} 
    & \sum_q\lambda_s^q(-2 e_1-2 e_2+4 e_5 \\ 
    & +2 e_7-e_8-e_9+2 e_{10} \\ 
    & -e_{11}-e_{12}+2 e_{13}+2 e_{14}-e_{18}) 
    \end{aligned}$ \tabularnewline\hline
    
    $\Lambda_b^0\to \Xi^- \pi^+  K^0 K^0$ & 
    $\begin{aligned} 
    & \sum_q\lambda_d^q(-e_3-e_4+2 e_6 \\ 
    & -e_7-e_8+2 e_9-e_{10} \\ 
    & -e_{11}+2 e_{12}+e_{15}+e_{16}+e_{18}) 
    \end{aligned}$ &
    $\Xi_b^0\to \Sigma^- K^+  \bar{K}^0 \bar{K}^0$ & 
    $\begin{aligned} 
    & \sum_q\lambda_s^q(e_3+e_4-2 e_6 \\ 
    & +e_7+e_8-2 e_9+e_{10} \\ 
    & +e_{11}-2 e_{12}-e_{15}-e_{16}-e_{18}) 
    \end{aligned}$ \tabularnewline\hline
    
    $\Xi_b^0\to \Xi^- K^+  \bar{K}^0 K^0$ & 
    $\begin{aligned} 
    & \sum_q\lambda_d^q(-2 e_1+4 e_2+2 e_3 \\ 
    & -e_4-2 e_5-e_6-e_{10} \\ 
    & +2 e_{11}-e_{12}-2 e_{13}-e_{16}+2 e_{17}) 
    \end{aligned}$ &
    $\Lambda_b^0\to \Sigma^- \pi^+  K^0 \bar{K}^0$ & 
    $\begin{aligned} 
    & \sum_q\lambda_s^q(2 e_1-4 e_2-2 e_3 \\ 
    & +e_4+2 e_5+e_6+e_{10} \\ 
    & -2 e_{11}+e_{12}+2 e_{13}+e_{16}-2 e_{17}) 
    \end{aligned}$ \tabularnewline\hline
\hline
    $\Lambda_b^0\to \Sigma^- \pi^+  K^+ \pi^-$ & 
    $\begin{aligned} 
    & \sum_q\lambda_d^q(-2 f_1+f_2-f_3+f_4 \\ 
    & -f_5+f_7+f_8+f_9 +f_{10}) 
    \end{aligned}$ &
    $\Xi_b^0\to \Xi^- K^+  \pi^+ K^-$ & 
    $\begin{aligned} 
    & \sum_q\lambda_s^q(2 f_1-f_2+f_3-f_4 \\ 
    & +f_5-f_7-f_8 -f_9-f_{10}) 
    \end{aligned}$ \tabularnewline\hline
  $\Lambda_b^0\to \Sigma^- K^+  K^+ K^-$ & $ \sum_q\lambda_d^q(-2 f_1+f_2+f_4+f_7+f_9)$ & $\Xi_b^0\to \Xi^- \pi^+  \pi^+ \pi^-$ & $ \sum_q\lambda_s^q(2 f_1-f_2-f_4-f_7-f_9)$ \tabularnewline\hline
  $\Xi_b^0\to \Sigma^- \pi^+  \pi^+ \pi^-$ & $ \sum_q\lambda_d^q(f_1+f_2+f_3-2 f_4-f_6-f_7-f_8)$ & $\Lambda_b^0\to \Xi^- K^+  K^+ K^-$ & $ \sum_q\lambda_s^q(-f_1-f_2-f_3+2 f_4+f_6+f_7+f_8)$ \tabularnewline\hline
  $\Xi_b^0\to \Sigma^- \pi^+  K^+ K^-$ & $ \sum_q\lambda_d^q(f_1+f_2-2 f_4-f_5-f_6-f_7+f_{10})$ & $\Lambda_b^0\to \Xi^- \pi^+  K^+ \pi^-$ & $ \sum_q\lambda_s^q(-f_1-f_2+2 f_4+f_5+f_6+f_7-f_{10})$ \tabularnewline\hline
    \hline
    \end{tabular}
    \end{table*}

    \begin{table*}[htbp!]
        \caption{The U-spin amplitudes of charmless bottomed baryon three body decays. Here $\lambda_s^u=V_{ub}^*V_{us}$, $\lambda_d^u=V_{ub}^*V_{ud}$, $\lambda_s^c=V_{cb}^*V_{cs}$ and $\lambda_d^c=V_{cb}^*V_{cd}$. The $\sum_q$ means sum the index $q$ with $q=u,c$.}
        \label{table3}
        \renewcommand{\arraystretch}{1.5}
        \begin{tabular}{cccc}
        \hline\hline
        Channel & Amplitude & Channel & Amplitude \\
        \hline 
        % A类配对
        $\Xi_{b}^{-}\to \Sigma^{0} K^{0}  \bar{K}^{0} \pi^{-} $ & 
        $\sum_q\lambda_{d}^{q}\big(\frac{1}{2\sqrt{2}}(A_2-A_1) +\frac{\sqrt{3}}{2}A^1_1\big)$ &
        $\Xi_{b}^{-}\to \Sigma^{0} K^{0}  \bar{K}^{0} K^{-} $ & 
        $
        \sum_q\lambda_{s}^{q}\big(\frac{1}{2\sqrt{2}}(A_1-A_2) 
        +\frac{\sqrt{3}}{2}A^1_1\big)
        $ \tabularnewline\hline
        
        $\Xi_{b}^{-}\to \Lambda^{0} K^{0}  \bar{K}^{0} \pi^{-} $ & 
        $
        \sum_q\lambda_{d}^{q}\big(-\frac{\sqrt3}{2\sqrt2}(A_2-A_1)  
        -\frac{1}{2}A^1_1\big)
        $ &
        $\Xi_{b}^{-}\to \Lambda^{0} K^{0}  \bar{K}^{0} K^{-} $ & 
        $
        \sum_q\lambda_{s}^{q}\big(\frac{\sqrt3}{2\sqrt2}(A_1-A_2)
        -\frac{1}{2}A^1_1\big)
        $ \tabularnewline\hline
        
        $\Xi_{b}^{-}\to n K^{0}  \bar{K}^{0} K^{-} $ & 
        $\sum_q\lambda_{d}^{q}(A_1+A_3)$ &
        $\Xi_{b}^{-}\to \Xi^{0} K^{0}  \bar{K}^{0} \pi^{-} $ & 
        $\sum_q\lambda_{s}^{q}(A_1+A_3)$ \tabularnewline\hline
        
        $\Xi_{b}^{-}\to \Xi^{0} K^{0}  K^{0} K^{-} $ & 
        $\sum_q\lambda_{d}^{q}(A_2+A_3)$ &
        $\Xi_{b}^{-}\to n \bar{K}^{0}  \bar{K}^{0} \pi^{-} $ & 
        $\sum_q\lambda_{s}^{q}(A_2+A_3)$ \tabularnewline\hline
        \hline
        % B类配对
        $\Xi_{b}^{-}\to \Sigma^{0} \pi^{-}  \pi^{-} \pi^{+} $ & 
        $
        \sum_q\lambda_{d}^{q}\big(-\frac{1}{2\sqrt{2}}(B_1+B_2)
        +\frac{\sqrt{3}}{2}B^1_1\big)
        $ &
        $\Xi_{b}^{-}\to \Sigma^{0} K^{-}  K^{-} K^{+} $ & 
        $
        \sum_q\lambda_{s}^{q}\big(\frac{1}{2\sqrt{2}}(B_1+B_2)
        +\frac{\sqrt{3}}{2}B^1_1\big)
        $ \tabularnewline\hline
        
        $\Xi_{b}^{-}\to \Lambda^{0} \pi^{-}  \pi^{-} \pi^{+} $ & 
        $
        \sum_q\lambda_{d}^{q}\big(-\frac{\sqrt3}{2\sqrt2}(B_1+B_2) 
        -\frac{1}{2}B^1_1\big)
        $ &
        $\Xi_{b}^{-}\to \Lambda^{0} K^{-}  K^{-} K^{+} $ & 
        $
        \sum_q\lambda_{s}^{q}\big(\frac{\sqrt3}{2\sqrt2}(B_1+B_2)  
        -\frac{1}{2}B^1_1\big)
        $ \tabularnewline\hline
        
        $\Xi_{b}^{-}\to \Sigma^{0} \pi^{-}  K^{-} K^{+} $ & 
        $
        \sum_q\lambda_{d}^{q}\big(\frac{1}{2\sqrt{2}}(B_2-B_1) 
        +\frac{\sqrt{3}}{2}B^1_1\big)
        $ &
        $\Xi_{b}^{-}\to \Sigma^{0} \pi^{-}  K^{-} \pi^{+} $ & 
        $
        \sum_q\lambda_{s}^{q}\big(\frac{1}{2\sqrt{2}}(B_1-B_2)  
        +\frac{\sqrt{3}}{2}B^1_1\big)
        $ \tabularnewline\hline
        
        $\Xi_{b}^{-}\to \Lambda^{0} \pi^{-}  K^{-} K^{+} $ & 
        $
        \sum_q\lambda_{d}^{q}\big(\frac{\sqrt3}{2\sqrt2}(B_2-B_1) 
        -\frac{1}{2}B^1_1\big)
        $ &
        $\Xi_{b}^{-}\to \Lambda^{0} \pi^{-}  K^{-} \pi^{+} $ & 
        $
        \sum_q\lambda_{s}^{q}\big(\frac{\sqrt3}{2\sqrt2}(B_1-B_2) 
        -\frac{1}{2}B^1_1\big)
        $ \tabularnewline\hline
        
        $\Xi_{b}^{-}\to n \pi^{-}  K^{-} \pi^{+} $ & 
        $\sum_q\lambda_{d}^{q}(B_1+B_2)$ &
        $\Xi_{b}^{-}\to \Xi^{0} \pi^{-}  K^{-} K^{+} $ & 
        $\sum_q\lambda_{s}^{q}(B_1+B_2)$ \tabularnewline\hline
        
        $\Xi_{b}^{-}\to n K^{-}  K^{-} K^{+} $ & 
        $\sum_q\lambda_{d}^{q}B_1$ &
        $\Xi_{b}^{-}\to \Xi^{0} \pi^{-}  \pi^{-} \pi^{+} $ & 
        $\sum_q\lambda_{s}^{q}B_1$ \tabularnewline\hline
        
        $\Xi_{b}^{-}\to \Xi^{0} \pi^{-}  \pi^{-} K^{+} $ & 
        $\sum_q\lambda_{d}^{q}B_2$ &
        $\Xi_{b}^{-}\to n K^{-}  K^{-} \pi^{+} $ & 
        $\sum_q\lambda_{s}^{q}B_2$ \tabularnewline\hline
        \hline
        % C类配对
        $\Xi_{b}^{-}\to p K^{0}  K^{-} K^{-} $ & 
        $\sum_q\lambda_{d}^{q}(C_1+C_3)$ &
        $\Xi_{b}^{-}\to \Sigma^{+} K^{0}  K^{-} K^{-} $ & 
        $\sum_q\lambda_{s}^{q}(C_1+C_2)$ \tabularnewline\hline
        
        $\Xi_{b}^{-}\to p \bar{K}^{0}  \pi^{-} \pi^{-} $ & 
        $-\sum_q\lambda_{d}^{q}(C_1+C_2)$ &
        $\Xi_{b}^{-}\to \Sigma^{+} \bar{K}^{0}  \pi^{-} \pi^{-} $ & 
        $-\sum_q\lambda_{s}^{q}(C_1+C_3)$ \tabularnewline\hline
        
        $\Xi_{b}^{-}\to \Sigma^{+} K^{0}  K^{-} \pi^{-} $ & 
        $\sum_q\lambda_{d}^{q}(C_2-C_3)$ &
        $\Xi_{b}^{-}\to p \bar{K}^{0}  K^{-} \pi^{-} $ & 
        $\sum_q\lambda_{s}^{q}(C_3-C_2)$ \tabularnewline\hline
        \hline
        % D类配对
        $\Xi_{b}^{-}\to \Sigma^{-} K^{0}  K^{-} \pi^{+} $ & 
        $\sum_q\lambda_{d}^{q}(D_1+D_4)$ &
        $\Xi_{b}^{-}\to \Xi^{-} \bar{K}^{0}  \pi^{-} K^{+} $ & 
        $\sum_q\lambda_{s}^{q}(D_1+D_4)$ \tabularnewline\hline
        
        $\Xi_{b}^{-}\to \Sigma^{-} \bar{K}^{0}  \pi^{-} K^{+} $ & 
        $\sum_q\lambda_{d}^{q}(D_1+D_3)$ &
        $\Xi_{b}^{-}\to \Xi^{-} K^{0}  K^{-} \pi^{+} $ & 
        $\sum_q\lambda_{s}^{q}(D_1+D_3)$ \tabularnewline\hline
        
        $\Xi_{b}^{-}\to \Xi^{-} K^{0}  \pi^{-} \pi^{+} $ & 
        $\sum_q\lambda_{d}^{q}(D_2+D_3)$ &
        $\Xi_{b}^{-}\to \Sigma^{-} \bar{K}^{0}  K^{-} K^{+} $ & 
        $\sum_q\lambda_{s}^{q}(D_2+D_3)$ \tabularnewline\hline
        
        $\Xi_{b}^{-}\to \Xi^{-} K^{0}  K^{-} K^{+} $ & 
        $\sum_q\lambda_{d}^{q}(D_2+D_4)$ &
        $\Xi_{b}^{-}\to \Sigma^{-} \bar{K}^{0}  \pi^{-} \pi^{+} $ & 
        $\sum_q\lambda_{s}^{q}(D_2+D_4)$ \tabularnewline\hline
        \hline
        % E类配对
        $\Xi_{b}^{-}\to \Xi^{-} K^{0}  \bar{K}^{0} K^{0} $ & 
        $\sum_q\lambda_{d}^{q}E_1$ &
        $\Xi_{b}^{-}\to \Sigma^{-} K^{0}  \bar{K}^{0} \bar{K}^{0} $ & 
        $\sum_q\lambda_{s}^{q}E_1$ \tabularnewline\hline
        \hline\hline
        \end{tabular}
\end{table*}
 In these tables, one can clearly observe that the amplitude relationships cannot be directly obtained, since the CKM matrix elements differ between $b\to d$ and $b\to s $ processes. To clearly describe the CPV relations, we divide the amplitude into two parts corresponding to its CKM matrix as: ${\cal A} = V_{ub}^* V_{uq} {\cal A}_u + V_{cb}^* V_{cq} {\cal A}_c$ and the relations of amplitude ${\cal A}_{u/c}$ can be directly derived. With the help of the definition of the ${\cal A}_{u/c}$ the CPV observable ${\cal A}^{dir}_{CP}$ can be expressed as
\begin{eqnarray}
A_{CP}^{dir}&=&-\frac{4{\cal I}m(V_{ub}^*V_{uq}V_{cb}V_{cq}^*){\cal I}m({\cal A}_u{\cal A}^*_c)}{|{\cal A}(B_b\to B MMM)|^2+|{\cal A}(\bar B_b\to \bar B \bar M\bar M\bar M)|^2},\quad q=d,s.
\end{eqnarray}
Since the CKM matrix unitarity requires ${\cal I}m(V_{ub}^*V_{us}V_{cb}V_{cs}^*)=-{\cal I}m(V_{ub}^*V_{ud}V_{cb}V^*_{cd})$, the CPV relations can be derived if the amplitudes ${\cal A}_{u/c}$ are directly related. These CPV relations can be expressed as
\begin{eqnarray}
A_{CP}^{dir}(B_b\to B MMM) &=&-A_{CP}^{dir}(B^\prime_b\to B^\prime M^\prime M^\prime M^\prime){\cal R}(\frac{B^\prime_b\to B^\prime M^\prime M^\prime M^\prime}{B_b\to B MMM}),
\end{eqnarray}
with
\begin{eqnarray}
{\cal R}(\frac{B_b\to B MMM}{B^\prime_b\to B^\prime M^\prime M^\prime M^\prime})=\frac{{\cal B}(B_b\to B MMM)\cdot \tau(B_b^\prime)}{{\cal B}(B^\prime_b\to B^\prime M^\prime M^\prime M^\prime)\cdot\tau(B_b)}.
\end{eqnarray}
The detailed derivation can be found in ref.~\cite{Wang:2024rwf}.

Using the same strategy, We can derive the CPV relations for b-baryon four body decays as follows:
    \begin{align}
        &\frac{A_{CP}^{dir}(\Xi_b^0\to nK^0  \bar{K}^{0} \bar{K}^0)}{A_{CP}^{dir}(\Lambda_b^0\to \Xi^0K^0  \bar{K}^0 K^0)} = -{\cal R}\left(\frac{\Lambda_b^0\to \Xi^0K^0  \bar{K}^0 K^0}{\Xi_b^0\to nK^0  \bar{K}^{0} \bar{K}^0}\right),
        \frac{A_{CP}^{dir}(\Xi_b^0\to \Sigma^0K^0  \bar{K}^0 \bar{K}^0)}{A_{CP}^{dir}(\Lambda_b^0\to \Sigma^0K^0  \bar{K}^0 K^0)} = -{\cal R}\left(\frac{\Lambda_b^0\to \Sigma^0K^0  \bar{K}^0 K^0}{\Xi_b^0\to \Sigma^0K^0  \bar{K}^0 \bar{K}^0}\right), \notag\\
        &\frac{A_{CP}^{dir}(\Xi_b^0\to \Lambda^0K^0  \bar{K}^0 \bar{K}^0)}{A_{CP}^{dir}(\Lambda_b^0\to \Lambda^0K^0  \bar{K}^0 K^0)} = -{\cal R}\left(\frac{\Lambda_b^0\to \Lambda^0K^0  \bar{K}^0 K^0}{\Xi_b^0\to \Lambda^0K^0  \bar{K}^0 \bar{K}^0}\right)
        \frac{A_{CP}^{dir}(\Xi_b^0\to \Xi^0 K^0  \bar{K}^0 K^0)}{A_{CP}^{dir}(\Lambda_b^0\to n K^0  \bar{K}^0 \bar{K}^0)} = -{\cal R}\left(\frac{\Lambda_b^0\to n K^0  \bar{K}^0 \bar{K}^0}{\Xi_b^0\to \Xi^0 K^0  \bar{K}^0 K^0}\right),\notag \\
        &\frac{A_{CP}^{dir}(\Lambda_b^0\to \Sigma^0 K^0  \pi^- \pi^+)}{A_{CP}^{dir}(\Xi_b^0\to \Sigma^0 \bar{K}^0  K^- K^+)} = -{\cal R}\left(\frac{\Xi_b^0\to \Sigma^0 \bar{K}^0  K^- K^+}{\Lambda_b^0\to \Sigma^0 K^0  \pi^- \pi^+}\right),
        \frac{A_{CP}^{dir}(\Lambda_b^0\to \Sigma^0 K^0  K^- K^+)}{A_{CP}^{dir}(\Xi_b^0\to \Sigma^0 \bar{K}^0  \pi^- \pi^+)} = -{\cal R}\left(\frac{\Xi_b^0\to \Sigma^0 \bar{K}^0  \pi^- \pi^+}{\Lambda_b^0\to \Sigma^0 K^0  K^- K^+}\right), \notag\\
        &\frac{A_{CP}^{dir}(\Xi_b^0\to \Sigma^0 K^0  K^- \pi^+)}{A_{CP}^{dir}(\Lambda_b^0\to \Sigma^0 \bar{K}^0  \pi^- K^+)} = -{\cal R}\left(\frac{\Lambda_b^0\to \Sigma^0 \bar{K}^0  \pi^- K^+}{\Xi_b^0\to \Sigma^0 K^0  K^- \pi^+}\right), 
        \frac{A_{CP}^{dir}(\Xi_b^0\to \Sigma^0 \bar{K}^0  \pi^- K^+)}{A_{CP}^{dir}(\Lambda_b^0\to \Sigma^0 K^0  K^- \pi^+)} = -{\cal R}\left(\frac{\Lambda_b^0\to \Sigma^0 K^0  K^- \pi^+}{\Xi_b^0\to \Sigma^0 \bar{K}^0  \pi^- K^+}\right), \notag\\
        &\frac{A_{CP}^{dir}(\Lambda_b^0\to \Lambda^0 K^0  \pi^- \pi^+)}{A_{CP}^{dir}(\Xi_b^0\to \Lambda^0 \bar{K}^0  K^- K^+)} = -{\cal R}\left(\frac{\Xi_b^0\to \Lambda^0 \bar{K}^0  K^- K^+}{\Lambda_b^0\to \Lambda^0 K^0  \pi^- \pi^+}\right), 
        \frac{A_{CP}^{dir}(\Lambda_b^0\to \Lambda^0 K^0  K^- K^+)}{A_{CP}^{dir}(\Xi_b^0\to \Lambda^0 \bar{K}^0  \pi^- \pi^+)} = -{\cal R}\left(\frac{\Xi_b^0\to \Lambda^0 \bar{K}^0  \pi^- \pi^+}{\Lambda_b^0\to \Lambda^0 K^0  K^- K^+}\right), \notag\\
        &\frac{A_{CP}^{dir}(\Xi_b^0\to \Lambda^0 K^0  K^- \pi^+)}{A_{CP}^{dir}(\Lambda_b^0\to \Lambda^0 \bar{K}^0  \pi^- K^+)} = -{\cal R}\left(\frac{\Lambda_b^0\to \Lambda^0 \bar{K}^0  \pi^- K^+}{\Xi_b^0\to \Lambda^0 K^0  K^- \pi^+}\right), 
        \frac{A_{CP}^{dir}(\Xi_b^0\to \Lambda^0 \bar{K}^0  \pi^- K^+)}{A_{CP}^{dir}(\Lambda_b^0\to \Lambda^0 K^0  K^- \pi^+)} = -{\cal R}\left(\frac{\Lambda_b^0\to \Lambda^0 K^0  K^- \pi^+}{\Xi_b^0\to \Lambda^0 \bar{K}^0  \pi^- K^+}\right), \notag\\
        &\frac{A_{CP}^{dir}(\Lambda_b^0\to \Xi^0 K^0  \pi^- K^+)}{A_{CP}^{dir}(\Xi_b^0\to n \bar{K}^0  K^- \pi^+)} = -{\cal R}\left(\frac{\Xi_b^0\to n \bar{K}^0  K^- \pi^+}{\Lambda_b^0\to \Xi^0 K^0  \pi^- K^+}\right), 
        \frac{A_{CP}^{dir}(\Xi_b^0\to \Xi^0 K^0  \pi^- \pi^+)}{A_{CP}^{dir}(\Lambda_b^0\to n \bar{K}^0  K^- K^+)} = -{\cal R}\left(\frac{\Lambda_b^0\to n \bar{K}^0  K^- K^+}{\Xi_b^0\to \Xi^0 K^0  \pi^- \pi^+}\right), \notag\\
        &\frac{A_{CP}^{dir}(\Xi_b^0\to \Xi^0 K^0  K^- K^+)}{A_{CP}^{dir}(\Lambda_b^0\to n \bar{K}^0  \pi^- \pi^+)} = -{\cal R}\left(\frac{\Lambda_b^0\to n \bar{K}^0  \pi^- \pi^+}{\Xi_b^0\to \Xi^0 K^0  K^- K^+}\right), 
        \frac{A_{CP}^{dir}(\Lambda_b^0\to n K^0  K^- \pi^+)}{A_{CP}^{dir}(\Xi_b^0\to \Xi^0 \bar{K}^0  \pi^- K^+)} = -{\cal R}\left(\frac{\Xi_b^0\to \Xi^0 \bar{K}^0  \pi^- K^+}{\Lambda_b^0\to n K^0  K^- \pi^+}\right), \notag\\
        &\frac{A_{CP}^{dir}(\Lambda_b^0\to n \bar{K}^0  \pi^- K^+)}{A_{CP}^{dir}(\Xi_b^0\to \Xi^0 K^0  K^- \pi^+)} = -{\cal R}\left(\frac{\Xi_b^0\to \Xi^0 K^0  K^- \pi^+}{\Lambda_b^0\to n \bar{K}^0  \pi^- K^+}\right), 
        \frac{A_{CP}^{dir}(\Xi_b^0\to n \bar{K}^0  \pi^- \pi^+)}{A_{CP}^{dir}(\Lambda_b^0\to \Xi^0 K^0  K^- K^+)} = -{\cal R}\left(\frac{\Lambda_b^0\to \Xi^0 K^0  K^- K^+}{\Xi_b^0\to n \bar{K}^0  \pi^- \pi^+}\right), \notag\\
        &\frac{A_{CP}^{dir}(\Xi_b^0\to n \bar{K}^0  K^- K^+)}{A_{CP}^{dir}(\Lambda_b^0\to \Xi^0 K^0  \pi^- \pi^+)} = -{\cal R}\left(\frac{\Lambda_b^0\to \Xi^0 K^0  \pi^- \pi^+}{\Xi_b^0\to n \bar{K}^0  K^- K^+}\right),
        \frac{A_{CP}^{dir}(\Lambda_b^0\to p \pi^-  \pi^- \pi^+)}{A_{CP}^{dir}(\Xi_b^0\to \Sigma^+ K^-  \pi^- \pi^+)} = -{\cal R}\left(\frac{\Xi_b^0\to \Sigma^+ K^-  \pi^- \pi^+}{\Lambda_b^0\to p \pi^-  \pi^- \pi^+}\right), \notag\\
        &\frac{A_{CP}^{dir}(\Lambda_b^0\to p K^-  \pi^- K^+)}{A_{CP}^{dir}(\Xi_b^0\to \Sigma^+ K^-  \pi^- \pi^+)} = -{\cal R}\left(\frac{\Xi_b^0\to \Sigma^+ K^-  \pi^- \pi^+}{\Lambda_b^0\to p K^-  \pi^- K^+}\right),
        \frac{A_{CP}^{dir}(\Xi_b^0\to p K^-  \pi^- \pi^+)}{A_{CP}^{dir}(\Lambda_b^0\to \Sigma^+ K^-  \pi^- K^+)} = -{\cal R}\left(\frac{\Lambda_b^0\to \Sigma^+ K^-  \pi^- K^+}{\Xi_b^0\to p K^-  \pi^- \pi^+}\right),   \notag\\
        &\frac{A_{CP}^{dir}(\Xi_b^0\to p K^-  K^- K^+)}{A_{CP}^{dir}(\Lambda_b^0\to \Sigma^+ \pi^-  \pi^- \pi^+)} = -{\cal R}\left(\frac{\Lambda_b^0\to \Sigma^+ \pi^-  \pi^- \pi^+}{\Xi_b^0\to p K^-  K^- K^+}\right), 
        \frac{A_{CP}^{dir}(\Lambda_b^0\to \Sigma^+ \pi^-  \pi^- K^+)}{A_{CP}^{dir}(\Xi_b^0\to p K^-  K^- \pi^+)} = -{\cal R}\left(\frac{\Xi_b^0\to p K^-  K^- \pi^+}{\Lambda_b^0\to \Sigma^+ \pi^-  \pi^- K^+}\right), \notag\\
        &\frac{A_{CP}^{dir}(\Xi_b^0\to \Sigma^+ \pi^-  \pi^- \pi^+)}{A_{CP}^{dir}(\Lambda_b^0\to p K^-  K^- K^+)} = -{\cal R}\left(\frac{\Lambda_b^0\to p K^-  K^- K^+}{\Xi_b^0\to \Sigma^+ \pi^-  \pi^- \pi^+}\right),
        \frac{A_{CP}^{dir}(\Xi_b^0\to \Sigma^+ \pi^-  K^- K^+)}{A_{CP}^{dir}(\Lambda_b^0\to p \pi^-  K^- \pi^+)} = -{\cal R}\left(\frac{\Lambda_b^0\to p \pi^-  K^- \pi^+}{\Xi_b^0\to \Sigma^+ \pi^-  K^- K^+}\right), \notag\\
        &\frac{A_{CP}^{dir}(\Lambda_b^0\to p K^0 \bar{K}^0 \pi^-)}{A_{CP}^{dir}(\Xi_b^0\to \Sigma^+ \bar{K}^0 K^0 K^-)} = -{\cal R}\left(\frac{\Xi_b^0\to \Sigma^+ \bar{K}^0 K^0 K^-}{\Lambda_b^0\to p K^0 \bar{K}^0 \pi^-}\right),
        \frac{A_{CP}^{dir}(\Xi_b^0\to p \bar{K}^0 K^0 K^-)}{A_{CP}^{dir}(\Lambda_b^0\to \Sigma^+ K^0 \bar{K}^0 \pi^-)} = -{\cal R}\left(\frac{\Lambda_b^0\to \Sigma^+ K^0 \bar{K}^0 \pi^-}{\Xi_b^0\to p \bar{K}^0 K^0 K^-}\right),  \notag\\
        &\frac{A_{CP}^{dir}(\Lambda_b^0\to \Sigma^+ K^0 K^0 K^-)}{A_{CP}^{dir}(\Xi_b^0\to p \bar{K}^0 \bar{K}^0 \pi^-)} = -{\cal R}\left(\frac{\Xi_b^0\to p \bar{K}^0 \bar{K}^0 \pi^-}{\Lambda_b^0\to \Sigma^+ K^0 K^0 K^-}\right),
        \frac{A_{CP}^{dir}(\Xi_b^0\to \Sigma^+ K^0 \bar{K}^0 \pi^-)}{A_{CP}^{dir}(\Lambda_b^0\to p \bar{K}^0 K^0 K^-)} = -{\cal R}\left(\frac{\Lambda_b^0\to p \bar{K}^0 K^0 K^-}{\Xi_b^0\to \Sigma^+ K^0 \bar{K}^0 \pi^-}\right),  \notag\\
        &\frac{A_{CP}^{dir}(\Lambda_b^0\to \Sigma^- K^+  K^0 \bar{K}^0)}{A_{CP}^{dir}(\Xi_b^0\to \Xi^- \pi^+  \bar{K}^0 K^0)} = -{\cal R}\left(\frac{\Xi_b^0\to \Xi^- \pi^+  \bar{K}^0 K^0}{\Lambda_b^0\to \Sigma^- K^+  K^0 \bar{K}^0}\right),
        \frac{A_{CP}^{dir}(\Xi_b^0\to \Sigma^- \pi^+  \bar{K}^0 K^0)}{A_{CP}^{dir}(\Lambda_b^0\to \Xi^- K^+  K^0 \bar{K}^0)} = -{\cal R}\left(\frac{\Lambda_b^0\to \Xi^- K^+  K^0 \bar{K}^0}{\Xi_b^0\to \Sigma^- \pi^+  \bar{K}^0 K^0}\right),  \notag\\
        &\frac{A_{CP}^{dir}(\Lambda_b^0\to \Xi^- \pi^+  K^0 K^0)}{A_{CP}^{dir}(\Xi_b^0\to \Sigma^- K^+  \bar{K}^0 \bar{K}^0)} = -{\cal R}\left(\frac{\Xi_b^0\to \Sigma^- K^+  \bar{K}^0 \bar{K}^0}{\Lambda_b^0\to \Xi^- \pi^+  K^0 K^0}\right),
        \frac{A_{CP}^{dir}(\Xi_b^0\to \Xi^- K^+  \bar{K}^0 K^0)}{A_{CP}^{dir}(\Lambda_b^0\to \Sigma^- \pi^+  K^0 \bar{K}^0)} = -{\cal R}\left(\frac{\Lambda_b^0\to \Sigma^- \pi^+  K^0 \bar{K}^0}{\Xi_b^0\to \Xi^- K^+  \bar{K}^0 K^0}\right), \notag\\ 
        &\frac{A_{CP}^{dir}(\Lambda_b^0\to \Sigma^- \pi^+  K^+ \pi^-)}{A_{CP}^{dir}(\Xi_b^0\to \Xi^- K^+  \pi^+ K^-)} = -{\cal R}\left(\frac{\Xi_b^0\to \Xi^- K^+  \pi^+ K^-}{\Lambda_b^0\to \Sigma^- \pi^+  K^+ \pi^-}\right),
        \frac{A_{CP}^{dir}(\Lambda_b^0\to \Sigma^- K^+  K^+ K^-)}{A_{CP}^{dir}(\Xi_b^0\to \Xi^- \pi^+  \pi^+ \pi^-)} = -{\cal R}\left(\frac{\Xi_b^0\to \Xi^- \pi^+  \pi^+ \pi^-}{\Lambda_b^0\to \Sigma^- K^+  K^+ K^-}\right), \notag\\
        &\frac{A_{CP}^{dir}(\Xi_b^0\to \Sigma^- \pi^+  \pi^+ \pi^-)}{A_{CP}^{dir}(\Lambda_b^0\to \Xi^- K^+  K^+ K^-)} = -{\cal R}\left(\frac{\Lambda_b^0\to \Xi^- K^+  K^+ K^-}{\Xi_b^0\to \Sigma^- \pi^+  \pi^+ \pi^-}\right),
        \frac{A_{CP}^{dir}(\Xi_b^0\to \Sigma^- \pi^+  K^+ K^-)}{A_{CP}^{dir}(\Lambda_b^0\to \Xi^- \pi^+  K^+ \pi^-)} = -{\cal R}\left(\frac{\Lambda_b^0\to \Xi^- \pi^+  K^+ \pi^-}{\Xi_b^0\to \Sigma^- \pi^+  K^+ K^-}\right),  \notag\\
        &\frac{A_{CP}^{dir}(\Xi_b^{-}\to \Sigma^{0} K^{0}  \bar{K}^{0} \pi^{-})}{A_{CP}^{dir}(\Xi_b^{-}\to \Sigma^{0} K^{0}  \bar{K}^{0} K^{-})} = -{\cal R}\left(\frac{\Xi_b^{-}\to \Sigma^{0} K^{0}  \bar{K}^{0} K^{-}}{\Xi_b^{-}\to \Sigma^{0} K^{0}  \bar{K}^{0} \pi^{-}}\right),
        \frac{A_{CP}^{dir}(\Xi_b^{-}\to \Lambda^{0} K^{0}  \bar{K}^{0} \pi^{-})}{A_{CP}^{dir}(\Xi_b^{-}\to \Lambda^{0} K^{0}  \bar{K}^{0} K^{-})} = -{\cal R}\left(\frac{\Xi_b^{-}\to \Lambda^{0} K^{0}  \bar{K}^{0} K^{-}}{\Xi_b^{-}\to \Lambda^{0} K^{0}  \bar{K}^{0} \pi^{-}}\right), \notag\\
        &\frac{A_{CP}^{dir}(\Xi_b^{-}\to n K^{0}  \bar{K}^{0} K^{-})}{A_{CP}^{dir}(\Xi_b^{-}\to \Xi^{0} K^{0}  \bar{K}^{0} \pi^{-})} = -{\cal R}\left(\frac{\Xi_b^{-}\to \Xi^{0} K^{0}  \bar{K}^{0} \pi^{-}}{\Xi_b^{-}\to n K^{0}  \bar{K}^{0} K^{-}}\right),
        \frac{A_{CP}^{dir}(\Xi_b^{-}\to \Xi^{0} K^{0}  K^{0} K^{-})}{A_{CP}^{dir}(\Xi_b^{-}\to n \bar{K}^{0}  \bar{K}^{0} \pi^{-})} = -{\cal R}\left(\frac{\Xi_b^{-}\to n \bar{K}^{0}  \bar{K}^{0} \pi^{-}}{\Xi_b^{-}\to \Xi^{0} K^{0}  K^{0} K^{-}}\right),  \notag\\
        &\frac{A_{CP}^{dir}(\Xi_b^{-}\to \Sigma^{0} \pi^{-}  \pi^{-} \pi^{+})}{A_{CP}^{dir}(\Xi_b^{-}\to \Sigma^{0} K^{-}  K^{-} K^{+})} = -{\cal R}\left(\frac{\Xi_b^{-}\to \Sigma^{0} K^{-}  K^{-} K^{+}}{\Xi_b^{-}\to \Sigma^{0} \pi^{-}  \pi^{-} \pi^{+}}\right),
        \frac{A_{CP}^{dir}(\Xi_b^{-}\to \Lambda^{0} \pi^{-}  \pi^{-} \pi^{+})}{A_{CP}^{dir}(\Xi_b^{-}\to \Lambda^{0} K^{-}  K^{-} K^{+})} = -{\cal R}\left(\frac{\Xi_b^{-}\to \Lambda^{0} K^{-}  K^{-} K^{+}}{\Xi_b^{-}\to \Lambda^{0} \pi^{-}  \pi^{-} \pi^{+}}\right), \notag\\
        &\frac{A_{CP}^{dir}(\Xi_b^{-}\to \Sigma^{0} \pi^{-}  K^{-} K^{+})}{A_{CP}^{dir}(\Xi_b^{-}\to \Sigma^{0} \pi^{-}  K^{-} \pi^{+})} = -{\cal R}\left(\frac{\Xi_b^{-}\to \Sigma^{0} \pi^{-}  K^{-} \pi^{+}}{\Xi_b^{-}\to \Sigma^{0} \pi^{-}  K^{-} K^{+}}\right),
        \frac{A_{CP}^{dir}(\Xi_b^{-}\to \Lambda^{0} \pi^{-}  K^{-} K^{+})}{A_{CP}^{dir}(\Xi_b^{-}\to \Lambda^{0} \pi^{-}  K^{-} \pi^{+})} = -{\cal R}\left(\frac{\Xi_b^{-}\to \Lambda^{0} \pi^{-}  K^{-} \pi^{+}}{\Xi_b^{-}\to \Lambda^{0} \pi^{-}  K^{-} K^{+}}\right), \notag\\
        &\frac{A_{CP}^{dir}(\Xi_b^{-}\to n \pi^{-}  K^{-} \pi^{+})}{A_{CP}^{dir}(\Xi_b^{-}\to \Xi^{0} \pi^{-}  K^{-} K^{+})} = -{\cal R}\left(\frac{\Xi_b^{-}\to \Xi^{0} \pi^{-}  K^{-} K^{+}}{\Xi_b^{-}\to n \pi^{-}  K^{-} \pi^{+}}\right),
        \frac{A_{CP}^{dir}(\Xi_b^{-}\to n K^{-}  K^{-} K^{+})}{A_{CP}^{dir}(\Xi_b^{-}\to \Xi^{0} \pi^{-}  \pi^{-} \pi^{+})} = -{\cal R}\left(\frac{\Xi_b^{-}\to \Xi^{0} \pi^{-}  \pi^{-} \pi^{+}}{\Xi_b^{-}\to n K^{-}  K^{-} K^{+}}\right),  \notag\\
        &\frac{A_{CP}^{dir}(\Xi_b^{-}\to \Xi^{0} \pi^{-}  \pi^{-} K^{+})}{A_{CP}^{dir}(\Xi_b^{-}\to n K^{-}  K^{-} \pi^{+})} = -{\cal R}\left(\frac{\Xi_b^{-}\to n K^{-}  K^{-} \pi^{+}}{\Xi_b^{-}\to \Xi^{0} \pi^{-}  \pi^{-} K^{+}}\right),
        \frac{A_{CP}^{dir}(\Xi_b^{-}\to p K^{0}  K^{-} K^{-})}{A_{CP}^{dir}(\Xi_b^{-}\to \Sigma^{+} K^{0}  K^{-} K^{-})} = -{\cal R}\left(\frac{\Xi_b^{-}\to \Sigma^{+} K^{0}  K^{-} K^{-}}{\Xi_b^{-}\to p K^{0}  K^{-} K^{-}}\right),  \notag\\
        &\frac{A_{CP}^{dir}(\Xi_b^{-}\to p \bar{K}^{0}  \pi^{-} \pi^{-})}{A_{CP}^{dir}(\Xi_b^{-}\to \Sigma^{+} \bar{K}^{0}  \pi^{-} \pi^{-})} = -{\cal R}\left(\frac{\Xi_b^{-}\to \Sigma^{+} \bar{K}^{0}  \pi^{-} \pi^{-}}{\Xi_b^{-}\to p \bar{K}^{0}  \pi^{-} \pi^{-}}\right),
        \frac{A_{CP}^{dir}(\Xi_b^{-}\to \Sigma^{+} K^{0}  K^{-} \pi^{-})}{A_{CP}^{dir}(\Xi_b^{-}\to p \bar{K}^{0}  K^{-} \pi^{-})} = -{\cal R}\left(\frac{\Xi_b^{-}\to p \bar{K}^{0}  K^{-} \pi^{-}}{\Xi_b^{-}\to \Sigma^{+} K^{0}  K^{-} \pi^{-}}\right), \notag\\
        &\frac{A_{CP}^{dir}(\Xi_b^{-}\to \Sigma^{-} K^{0}  K^{-} \pi^{+})}{A_{CP}^{dir}(\Xi_b^{-}\to \Xi^{-} \bar{K}^{0}  \pi^{-} K^{+})} = -{\cal R}\left(\frac{\Xi_b^{-}\to \Xi^{-} \bar{K}^{0}  \pi^{-} K^{+}}{\Xi_b^{-}\to \Sigma^{-} K^{0}  K^{-} \pi^{+}}\right),
        \frac{A_{CP}^{dir}(\Xi_b^{-}\to \Sigma^{-} \bar{K}^{0}  \pi^{-} K^{+})}{A_{CP}^{dir}(\Xi_b^{-}\to \Xi^{-} K^{0}  K^{-} \pi^{+})} = -{\cal R}\left(\frac{\Xi_b^{-}\to \Xi^{-} K^{0}  K^{-} \pi^{+}}{\Xi_b^{-}\to \Sigma^{-} \bar{K}^{0}  \pi^{-} K^{+}}\right),  \notag\\
        &\frac{A_{CP}^{dir}(\Xi_b^{-}\to \Xi^{-} K^{0}  \pi^{-} \pi^{+})}{A_{CP}^{dir}(\Xi_b^{-}\to \Sigma^{-} \bar{K}^{0}  K^{-} K^{+})} = -{\cal R}\left(\frac{\Xi_b^{-}\to \Sigma^{-} \bar{K}^{0}  K^{-} K^{+}}{\Xi_b^{-}\to \Xi^{-} K^{0}  \pi^{-} \pi^{+}}\right),
        \frac{A_{CP}^{dir}(\Xi_b^{-}\to \Xi^{-} K^{0}  K^{-} K^{+})}{A_{CP}^{dir}(\Xi_b^{-}\to \Sigma^{-} \bar{K}^{0}  \pi^{-} \pi^{+})} = -{\cal R}\left(\frac{\Xi_b^{-}\to \Sigma^{-} \bar{K}^{0}  \pi^{-} \pi^{+}}{\Xi_b^{-}\to \Xi^{-} K^{0}  K^{-} K^{+}}\right),  \notag\\
        &\frac{A_{CP}^{dir}(\Xi_b^{-}\to \Xi^{-} K^{0}  \bar{K}^{0} K^{0})}{A_{CP}^{dir}(\Xi_b^{-}\to \Sigma^{-} K^{0}  \bar{K}^{0} \bar{K}^{0})} = -{\cal R}\left(\frac{\Xi_b^{-}\to \Sigma^{-} K^{0}  \bar{K}^{0} \bar{K}^{0}}{\Xi_b^{-}\to \Xi^{-} K^{0}  \bar{K}^{0} K^{0}}\right).\label{r}
\end{align}
One can see that with so many amplitudes in Table.\ref{table1}, Table.\ref{table2} and Table.\ref{table3}, huge CPV relations still can be derived. We also find that the relations of b-baryon U-spin doublet four body decays can only given the relation between $\Lambda_b^0$ and $\Xi_b^0$ as $\frac{A_{CP}^{dir}(\Lambda^0_b\to B MMM) }{A_{CP}^{dir}(\Xi^0b\to B^\prime M^\prime M^\prime M^\prime)}$. It means we can derive the CPV of $\Xi^0_b$ decays by the current experimental data. Using the experimental data~\cite{LHCb:2025ray}
\begin{eqnarray}
{\cal A}^{dir}_{CP}(\Lambda_b^0\to p \pi^-\pi^+K^-)=(2.45\pm0.46\pm 0.1)\%,
\end{eqnarray}
and~\cite{HeavyFlavorAveragingGroupHFLAV:2024ctg}
\begin{eqnarray}
\tau(\Lambda^0_b)=1.468\pm0.009 ps, \tau(\Xi_b^0)=1.477\pm0.032 ps,\tau(\Xi^-_b)=1.578\pm0.021 ps,
\end{eqnarray}
we can have
\begin{eqnarray}
A_{CP}^{dir}(\Xi_b^0 \to \Sigma^+ \pi^- K^- K^+) &=& -\mathcal{R}\left(\frac{\Lambda_b^0 \to p \pi^- K^- \pi^+}{\Xi_b^0 \to \Sigma^+ \pi^- K^- K^+}\right) \times A_{CP}^{dir}(\Lambda_b^0 \to p \pi^- K^- \pi^+)\notag\\
&=&-(2.45\pm0.46) \times \frac{(5.1\pm0.5)\times 10^{-5 }}{Br(\Xi_b^0 \to \Sigma^+ \pi^- K^- K^+)}\frac{\tau(\Xi_b)}{\tau(\Lambda_b)}=\frac{(-1.23\pm0.26)\times 10^{-4}}{Br(\Xi_b^0 \to \Sigma^+ \pi^- K^- K^+)}.
\end{eqnarray}
We notice that our result is consistent with other work~\cite{He:2025msg}. Without assumption, the U-spin analysis can only give one prediction with current data.
However, since the generation efficiency of $\Xi^0_b$ is not as high as that of $\Lambda^0_b$ in LHCb experiment, the prediction of $\Xi^0_b$ decay CPV may not be immediately useful for experimental CPV detection. Once one experimental data of CPV in Eq.\eqref{r} are measured in the future, the CPV of corresponding processes can be predicted if the U-spin symmetry is strictly conserved.

However, due to the mass difference between the $d$ quark and $s$ quark, U-spin symmetry is expected to be broken to some extent. Following the  Eq.\eqref{r}, we can define the deviation $\Delta$ to express the size of symmetry breaking.  The $\Delta$ is defined as
\begin{eqnarray}
\Delta=\frac{A_{CP}^{dir}(\Xi_b^0 \to \Sigma^+ \pi^- K^- K^+)}{A_{CP}^{dir}(\Lambda_b^0 \to p \pi^- K^- \pi^+)}+\mathcal{R}\left(\frac{\Lambda_b^0 \to p \pi^- K^- \pi^+}{\Xi_b^0 \to \Sigma^+ \pi^- K^- K^+}\right).
\end{eqnarray}
Once the CPV of both two processes in one U-spin relations are measured, the U-spin symmetry can be tested by the $\Delta$. Since the CPV of baryon is measured in the first time, the value of $\Delta$ can not be calculated in our work. 
We hope that sufficient experimental data will become available in the future, allowing us to determine the numerical value of $\Delta$ and thereby quantify the extent of U-spin symmetry breaking in b-baryon decays. 
Nevertheless, in our work we can still give a rough estimate of U-spin symmetry breaking. Since the mass difference between the down and strange quarks is small compared to that of the bottom quark, the breaking of U-spin symmetry is expected to be at the level of approximately $10\sim20\%$. This expectation is well supported by theoretical studies of charmless b-meson decays into two pseudoscalar mesons based on U-spin symmetry, where comparisons between theoretical predictions and experimental data show good agreement~\cite{He:2013vta}.

Furthermore, we observe that unlike the SU(3) method, the u-quark indices are not explicitly included in Hamiltonian matrix. Consequently, the color symmetry cannot be reflected by the decomposition of Hamiltonian. We can not omit any U-spin parameters using color symmetry, as is possible in SU(3) analyses. Therefore, in this step, we can provide one prediction with the only U-spin analysis. However, considering the complexity of four body decay, the decay structure analysis involving dynamic information will help simplify the U-spin amplitude. This approach will be given in the following section.

%%%%%%%%%%%%%%%%%%%%%%%%
\section{ simple dynamic analysis under U-spin analysis}
%%%%%%%%%%%%%%%%%%%%%%%%

For searching more useful prediction, we can give a deep analysis which involved dynamic information on the b-baryon doublet decays: $B_b^2\to B_1^2 M^2 M^2 M^{\bar 2}$. Its amplitude can be expressed as
\begin{eqnarray}
{\cal A}_{222 \bar2}^{2} &=& \sum_{q=u,c} \bigg( c_{1q} (B_{b}^2)^i (H_q)^j (B^2_1)_{i} (M^{2})_{j} (M^{2})_{k}(M^{\bar{2}})^k
+ c_{2q} (B_{b}^2)^j (H_q)^i (B^2_1)_{i} (M^{2})_{j} (M^{2})_{k}(M^{\bar{2}})^k \notag \\
&&\quad  \quad  \quad + c_{3q} (B_{b}^2)^j (H_q)^k (B^2_1)_{i} (M^{2})_{j} (M^{2})_{k}(M^{\bar{2}})^i \bigg), \label{am22222}
\end{eqnarray}
Although these amplitudes are given by the IRA method, fortunately, for these processes, the indices of matrix also reflect the quark d,s quark following the same method in SU(3) analysis\cite{He:2018php}. However, due to the absence of the u quark index, a single amplitude can correspond to multiple topological diagram. If we want to draw the u-quark line in topological diagram, analysis incorporating dynamic information becomes necessary. 

Besides, in LHCb experiment measurement, the CPV with different decay structure are measured as
\begin{eqnarray}
{\cal A}^{dir}_{CP}(\Lambda^0_b\to R(p K^-) R(\pi^+\pi^-))&=&(5.3\pm1.3\pm0.2)\%, \quad m_{p K^-}<2.2{\rm GeV}, \quad m_{\pi^-\pi^+}<1.1{\rm GeV},\notag\\
{\cal A}^{dir}_{CP}(\Lambda^0_b\to R(p \pi^-) R(\pi^+K^-))&=&(2.7\pm0.8\pm0.1)\%,\quad m_{p\pi^-}<1.7{\rm GeV},\notag\\
&&0.8{\rm GeV}<m_{\pi^+K^-}<1.0{\rm GeV}\; {\rm or}\; 1.1{\rm GeV}<m_{\pi^+K^-}<1.6{\rm GeV},\notag\\
{\cal A}^{dir}_{CP}(\Lambda^0_b\to R(p\pi^+\pi^-)K^-)&=&(5.4\pm0.9\pm0.1)\%,\quad m_{p\pi^+\pi^-}<2.7{\rm GeV},\notag\\
{\cal A}^{dir}_{CP}(\Lambda^0_b\to R( K^-\pi^+\pi^-)p)&=&(2.0\pm1.2\pm0.3)\%,\quad m_{K^-\pi^+\pi^-}<2.0 {\rm GeV}.\label{cpv}
\end{eqnarray}
It can be observed that the CPV is mainly contributed by the processes $\Lambda^0_b\to R(p\pi^+\pi^-)K^-$. To obtain the topological diagram of the  U-spin amplitude, we give a simple dynamic analysis.  

Generally, the b-baryon decay processes can be factorized into hard scattering kernel and wave function as
\begin{eqnarray}
{\cal A}(B_b\to B M_1 M_2 M_3)=h\otimes \phi_{B_b}\otimes \phi_{B}\otimes \phi_{M_1}\otimes \phi_{M_2}\otimes \phi_{M_3}.
\end{eqnarray}
The simple dynamic analysis can be given in Soft-Collinear Effective Theory(SCET) which has been applied in $\Lambda_b$ decay processes in 2012~\cite{Wang:2011uv}.
Although our analysis are not the strictly analysis under SCET, with the langrage of SCET, we can clearly show our logic and rationality of our analysis. 
For convenience, we can set the power of the large energy scale $Q\sim 1$. The momentum in the final state can be divided into four types: hard,  hard-collinear, collinear, and soft.  After integrating the hard mode in SCET$_I$  and the hard-collinear mode in SCET$_{II}$, the field in SCET can be divided into soft and collinear modes.  Under the light cone frame $p^\mu=(p^+,p_\perp,p^-)$, the momentum in collinear mode can be described as: $p_c^\mu\sim(1,\lambda^2,\lambda^4)$ with $\lambda=\sqrt{\frac{\Lambda_{QCD}}{m_b}}$. The soft mode is shown as $p_s^\mu\sim(\lambda^2,\lambda^2,\lambda^2)$.  With the help of definition of soft and collinear mode momentum, we can easily counting the power of the quark field $\psi=\psi_c+\psi_s$  as
\begin{eqnarray}
\xi_c=\frac{\slashed n_-\slashed n_+}{4}\psi_c\sim \lambda^2,\quad h_v=\frac{1+\slashed v}{2}Q_v\sim \lambda^3,\quad q_s\sim \lambda^3,
\end{eqnarray}
where $Q_v$ is the heavy quark field in Heavy Quark Effective Theory(HQET). The definition of vector $n^\mu_-$, $n^\mu_+$ and $v^\mu$ can be found in Ref.\cite{Beneke:2002ph}. Then the power of the Hamiltonian and the matrix element corresponding to different topological diagrams can be counted.

\begin{figure}[htbp!]
  \centering
\includegraphics[width=1\linewidth]{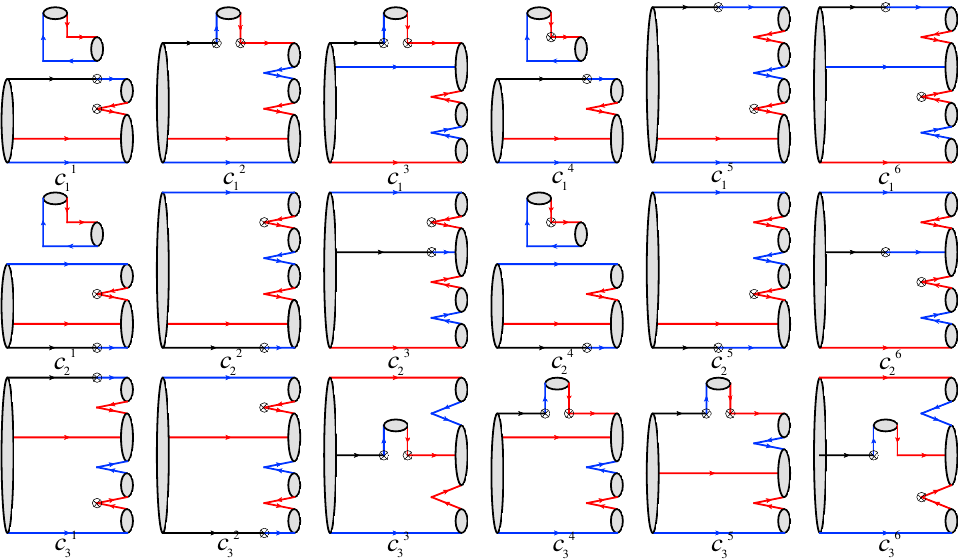} 
\caption{The leading power topological diagram of $B_b^2\to B_1^2 M^2 M^2 M^{\bar 2}$. The blue line indicates the d and s quark/anti-quark, the black line indicates the b quark. The red line indicates the u quark and u anti-quark. }
\label{tgc}
\end{figure}

As for the  Hamiltonian $[\bar Q \psi][\bar\psi \psi]$ we gived in Eq.\eqref{HH}, the leading power contribution can be simply estimated by the field with specific power as 
\begin{eqnarray}
[\bar h_v q_s][\bar \xi_c \xi_c]\sim\lambda^{10}, \qquad [\bar h_v \xi_c][\bar \xi_c q_s]\sim\lambda^{10},\qquad  [\bar h_v \xi_c][\bar q_s \xi_c]\sim\lambda^{10} . \label{hk}
\end{eqnarray}
The specific expression and power counting of four quark operator  can be seen in the Ref.\cite{Hill:2002vw}. The power counting of Hamiltonian shows that the three light quark field can only have two collinear field and the last one must be soft. Using the leading power Hamiltonian, we can also counting the power of matrix element $\langle B_1^2 M^2 M^2 M^{\bar 2}|{\mathcal H}| B_b^2|\rangle$ corresponding to different topological diagrams. By draw the u-quark line, the topological diagrams corresponding to parameter $c_1$. $c_2$ and $c_3$ are given in Fig.\ref{tgc}.

Before we counting the power of each  topological diagram, there are three point we must the state in the following:
\begin{enumerate}
 \item[A.] In the rest frame of the b-baryon, the HQET indicates that the momentum of the light quark in b-baryon should be soft. 
\item[B.] The strictly analysis of matrix element in SCET need consider the strictly matching between  four quark operator and  several SCET operators in which the hard and hard-collinear gluon contribution will be integrated and therefore the hard scattering kernel  will introduced. However, since the hard and hard-collinear gluon line can not be drawn in the topological diagram, we must omit the scattering kernel $h$ and only analyze the contribution of SCET operator in Eq.\eqref{hk}. That is the reason we called the dynamic analysis in our work is simply analysis.  
\item[C.] The quark and anti-quark pair which come from the vacuum in Fig.\ref{tgc} are seen as  soft quark and anti-quark produced form a soft gluon in our work.
\end{enumerate}
Basic on these conclusion, the power of each topological diagram can be determined in our work. 
 
\begin{figure}[htbp!]
  \centering
\includegraphics[width=0.8\linewidth]{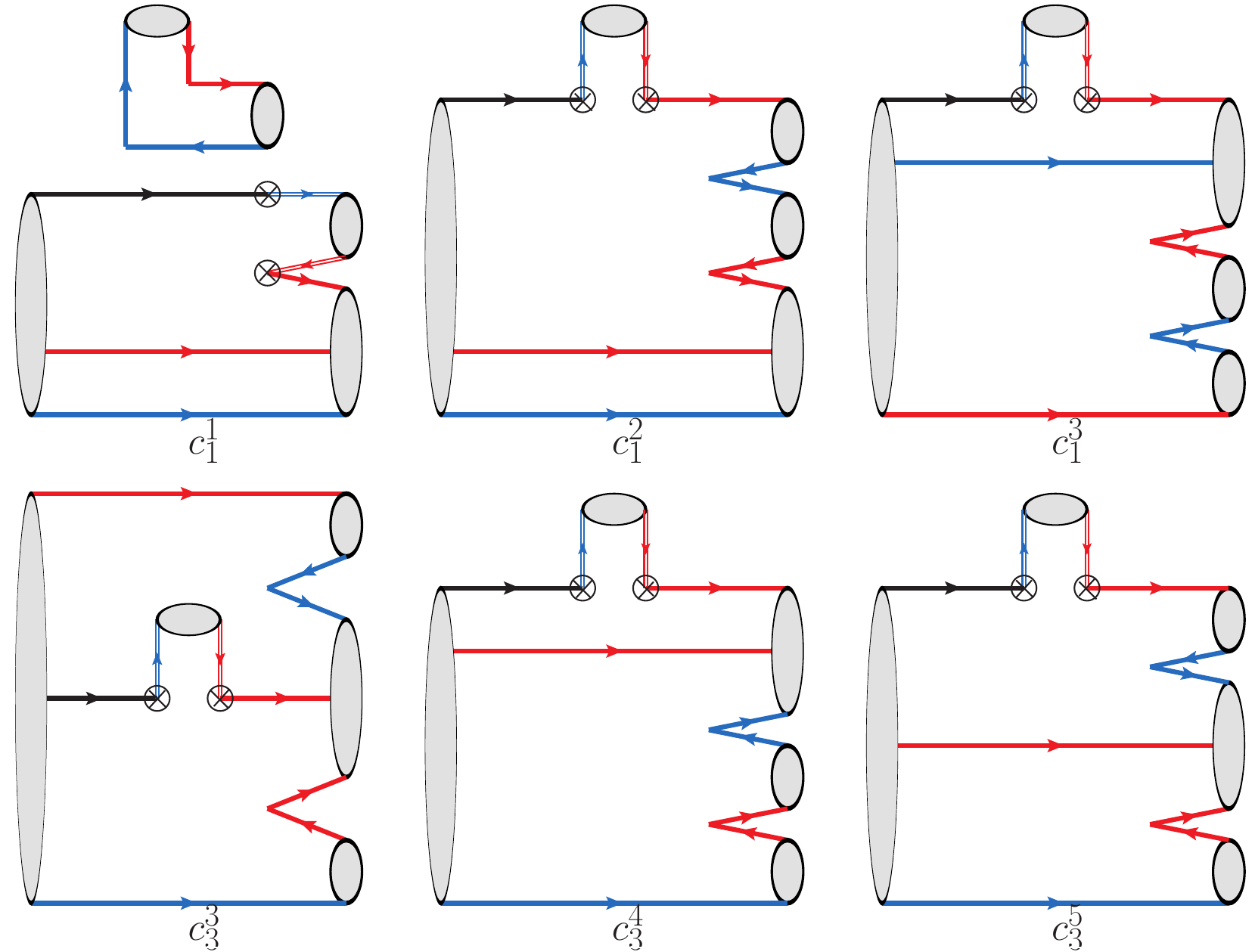} 
\caption{The leading power topological diagram of $B_b^2\to B_1^2 M^2 M^2 M^{\bar 2}$. The blue line indicates the d and s quark/anti-quark, the black line indicates the b quark. The red line indicates the u quark and u anti-quark. For expressing the dynamic information, we use the double lines to represent the collinear quark and the single to represent the soft quark. }
\label{6}
\end{figure}

Since the kinematic region of soft and collinear are different, the wave function of hadron containing both soft quark and collinear quark is power suppressed~\cite{Wang:2011uv}. Therefore the strategy of find the leading power topological diagram is choosing the suitable leading power Hamiltonian and making the quark field in light hadron are all soft or collinear. One can easily find that the leading power topological diagram  in Fig.\ref{tgc} are $c_1^1$, $c^2_1$, $c^3_1$, $c^3_3$, $c^4_3$ and $c^5_3$. The specific dynamic information of these diagram are shown in Fig.\ref{6}. 
 It indicates that there $c_1$ and $c_3$ can contribute the  $B_b^2\to B_1^2 M^2 M^2 M^{\bar 2}$ processes in the leading power. However, one can find that since the different decay structure are shown different in each topological diagram. The leading power topological diagram can not satnd for the total decay processes $ B_b^2\to B_1^2 M^2 M^2 M^{\bar 2}$ and the analysis focus on different decay structure must be given. 

In Fig.\ref{6}, one can find these topological diagram contain two structure. The first structure corresponding to $c_1^1$ is the soft baryon and collinear meson produced together and other two mesons produced together as $B_b^2\to R(B_1^2 M^2) R( M^{\bar 2} M^2)$. The another structure corresponding to $c_1^2,\;c_1^3,\;c_3^3,\;c_3^4,\;c_3^5$ is collinear meson emits alone and the soft meson and baryon produced together as $B_b^2\to R(B_1^2 M^2 M^{\bar 2}) M^2$.    Therefore if we focus on the structure $B_b^2\to R(B_1^2 M^2) R( M^{\bar 2} M^2)$, only amplitude $c_1$ have contribution and if we focus on the structure $B_b^2\to R(B_1^2 M^2 M^{\bar 2}) M^2$, only amplitudes $c_1$ and $c_3$ can have contributions.

%%%%%%%%%%%%%%%%%%%%%%%%
\section{ prediction and conclusion}
%%%%%%%%%%%%%%%%%%%%%%%%
Using the results of our simple analysis involve dynamic information, one can easily simplify the amplitude of $B_b^2\to B_1^2 M^2 M^2 M^{\bar 2}$. Since in Eq.\eqref{cpv} only the specific decay structure $\Lambda^0_b\to R(p\pi^+\pi^-)K^-$ and $\Lambda^0_b\to R(p K^-) R(\pi^+\pi^-)$ have observed the CPV, we will only focus on them in this section.
Considering this specific decay structure, the amplitude of $B_b^2\to R(B_1^2  M^2 M^{\bar 2})M^2$ and  $B_b^2\to R(B_1^2 M^2) R(  M^2 M^{\bar 2})$ can be given as
\begin{eqnarray}
 {\cal A}(\Lambda_b^0\to R(p \pi^-  \pi^+)\pi^-) &=&\sum_q\lambda_d^q(c_1+c_3), \quad {\cal A}(\Xi_b^0\to R(\Sigma^+ K^-  K^+) K^-)= \sum_q\lambda_s^q(c_1+c_3),\notag\\
 {\cal A}(\Xi_b^0\to R(\Sigma^+  K^- K^+)\pi^-) &=& \sum_q\lambda_d^q(c_1+c_3),\quad  {\cal A}(\Lambda_b^0\to R( p \pi^-  \pi^+)K^- )=\sum_q\lambda_s^q(c_1+c_3),\notag\\
{\cal A}(\Lambda_b^0\to R(p \pi^-) R(\pi^-  \pi^+))&=&\sum_q\lambda_d^qc_1,\quad {\cal A}(\Xi_b^0\to R(\Sigma^+ K^-)R( K^-  K^+) )= \sum_q\lambda_s^qc_1,\notag\\
{\cal A}(\Lambda_b^0\to R(p \pi^-) R(K^-  K^+))&=&\sum_q\lambda_d^qc_1,\quad {\cal A}(\Xi_b^0\to R(\Sigma^+ K^-)R( \pi^-  \pi^+) )= \sum_q\lambda_s^qc_1,\notag\\
{\cal A}(\Xi_b^0\to R(\Sigma^+ \pi^-) R(\pi^-  \pi^+))&=&\sum_q\lambda_d^qc_1, \quad {\cal A}(\Lambda_b^0\to R(p K^-)R( K^-  K^+) )= \sum_q\lambda_s^qc_1,\notag\\
{\cal A}(\Xi_b^0\to R(\Sigma^+ \pi^-) R( K^- K^+))&=& \sum_q\lambda_d^qc_1,\quad {\cal A}(\Lambda_b^0\to R( p K^-)( \pi^-  \pi^+) )=\sum_q\lambda_s^qc_1.\notag\\
\end{eqnarray}
Using these amplitude one can establish the CPV relations. Assuming the ratio of specific decay structure ${\cal R}(\frac{B_b\to R(B MM)M}{B^\prime_b\to R(B^\prime M^\prime M^\prime) M^\prime})$ and ${\cal R}(\frac{B_b\to R(B M)R(MM)}{B^\prime_b\to R(B^\prime M^\prime)R( M^\prime M^\prime)})$ are equal with full decay ratios ${\cal R}(\frac{B_b\to B MMM}{B^\prime_b\to B^\prime M^\prime M^\prime M^\prime})$, the relation can be written as
\begin{align}
    &\frac{A_{CP}^{dir}(\Lambda_b^0 \to R(p \pi^- \pi^+) \pi^-)}{A_{CP}^{dir}(\Lambda_b^0 \to R(p \pi^- \pi^+) K^-)} = -\mathcal{R}\left(\frac{\Lambda_b^0 \to p \pi^- K^- \pi^+}{\Lambda_b^0 \to p \pi^- \pi^- \pi^+}\right),\frac{A_{CP}^{dir}(\Xi_b^0 \to R(\Sigma^+ K^- K^+) \pi^-)}{A_{CP}^{dir}(\Lambda_b^0 \to R(p \pi^- \pi^+) K^-)} = -\mathcal{R}\left(\frac{\Lambda_b^0 \to p \pi^- K^- \pi^+}{\Xi_b^0 \to \Sigma^+ \pi^- K^- K^+}\right),\notag \\
    &\frac{A_{CP}^{dir}(\Xi_b^0\to R(\Sigma^+ K^- K^+)  K^-)}{A_{CP}^{dir}(\Lambda_b^0 \to R(p \pi^- \pi^+) K^-)} = \mathcal{R}\left(\frac{\Lambda_b^0 \to p \pi^- K^- \pi^+}{\Xi_b^0\to \Sigma^+ K^-  K^- K^+}\right), \frac{A_{CP}^{dir}(\Xi_b^0\to R(\Sigma^+ \pi^-)R( K^+  K^-))}{A_{CP}^{dir}(\Lambda_b^0 \to R(p  K^-)R(\pi^- \pi^+))} = -\mathcal{R}\left(\frac{\Lambda_b^0 \to p \pi^- K^- \pi^+}{\Xi_b^0\to \Sigma^+ \pi^-  K^- K^+}\right),\notag\\
    &\frac{A_{CP}^{dir}(\Lambda_b^0 \to R(p \pi^-)R( \pi^+ \pi^-))}{A_{CP}^{dir}(\Lambda_b^0 \to R(p K^-)R( \pi^- \pi^+))} = -\mathcal{R}\left(\frac{\Lambda_b^0 \to p \pi^- K^- \pi^+}{\Lambda_b^0 \to p \pi^- \pi^- \pi^+}\right),\frac{A_{CP}^{dir}(\Xi_b^0\to R(\Sigma^+ K^-)R( K^+  K^-))}{A_{CP}^{dir}(\Lambda_b^0 \to R(p  K^-)R(\pi^- \pi^+))} = \mathcal{R}\left(\frac{\Lambda_b^0 \to p \pi^- K^- \pi^+}{\Xi_b^0\to \Sigma^+ K^-  K^- K^+}\right), \notag\\
    &\frac{A_{CP}^{dir}(\Lambda_b^0 \to R(p \pi^-)R( K^+ K^-))}{A_{CP}^{dir}(\Lambda_b^0 \to R(p K^-)R( \pi^- \pi^+))} = -\mathcal{R}\left(\frac{\Lambda_b^0 \to p \pi^- K^- \pi^+}{\Lambda_b^0 \to p K^- \pi^- K^+}\right),\frac{A_{CP}^{dir}(\Xi_b^0\to R(\Sigma^+ K^-)R( \pi^+  \pi^-))}{A_{CP}^{dir}(\Lambda_b^0 \to R(p  K^-)R(\pi^- \pi^+))} = \mathcal{R}\left(\frac{\Lambda_b^0 \to p \pi^- K^- \pi^+}{\Xi_b^0\to \Sigma^+ \pi^-  K^- \pi^+}\right), \notag\\
    &\frac{A_{CP}^{dir}(\Xi_b^0 \to R(\Sigma^+ \pi^-)R( \pi^+ \pi^-))}{A_{CP}^{dir}(\Lambda_b^0 \to R(p K^-)R( \pi^- \pi^+))} = -\mathcal{R}\left(\frac{\Lambda_b^0 \to p \pi^- K^- \pi^+}{\Xi_b^0 \to \Sigma^+ \pi^- \pi^- \pi^+}\right),\frac{A_{CP}^{dir}(\Lambda_b^0\to R(p K^-)R( K^+  K^-))}{A_{CP}^{dir}(\Lambda_b^0 \to R(p  K^-)R(\pi^- \pi^+))} = \mathcal{R}\left(\frac{\Lambda_b^0 \to p \pi^- K^- \pi^+}{\Lambda_b^0\to p K^-  K^- K^+}\right).
\end{align}
For predicting the CPV, the experimental data for $\Lambda_b^0 \to p \pi^- \pi^- \pi^+$ ,  $\Lambda_b^0 \to p \pi^- K^- \pi^+$ and  $\Lambda_b^0 \to p \pi^- K^- K^+$ and  $\Lambda_b^0 \to p K^- K^- K^+$  decays are provided in~\cite{Rui:2023fiz}
\begin{align}
   & \mathcal{B} (\Lambda_b^0 \to p \pi^- \pi^- \pi^+) = (2.12 \pm0.21) \times 10^{-5},\quad \mathcal{B} (\Lambda_b^0 \to p \pi^- K^- \pi^+) = (5.1 \pm0.5) \times 10^{-5},\notag\\
   &\mathcal{B} (\Lambda_b^0 \to p \pi^- K^- K^+) = (4.1 \pm0.6) \times 10^{-6},\quad\mathcal{B} (\Lambda_b^0 \to p K^- K^- K^+) = (1.27 \pm0.13) \times 10^{-5}.
\end{align}
With these experimental data,  the U-spin symmetry can give many helpful predictions. However, there is one thing we need to clarify that the validity of our prediction are affected by the symmetry breaking effect. Therefore the error of our prediction can be estimated by the error of experimental data and symmetry breaking effect.  As for the symmetry breaking effect, constrained by the current data, the value of symmetry breaking effect can not be determined. Referenced by the previous research~\cite{He:2013vta}, the symmetry breaking effect can be estimated about $10\%\sim 20\%$. Therefore the error which comes from the symmetry breaking effect can be estimated by vary the value of prediction about $20\%$. Besides, the power expansion of our dynamic analysis also can affect the validity of our prediction with $\lambda\sim \sqrt{\frac{\Lambda_{QCD}}{m_b}}$.  Analyzing in the Ref.~\cite{Wang:2011uv}, the hadron wave function which contain soft and collinear quark will  leads to the phase suppression.  The meson wave function with soft and collinear quark will suppressed by $\lambda^2$ and the baryon wave function with two soft and one collinear quark will suppressed by $\lambda^4$. Therefore the contribution of next to leading power can be estimated by the Fig.\ref{tgc}  as $\lambda^4\sim5\%$.  
Then the CPV can be predicted as
\begin{align}
    &A_{CP}^{dir}(\Lambda_b^0 \to R( p \pi^- \pi^+ )\pi^-) = (-12.99 \pm 2.83_{\rm exp}\pm2.59_{\rm U-spin}\pm0.65_{\rm power})\%,\notag\\
    &A_{CP}^{dir}(\Lambda_b^0 \to R( p \pi^-)R( \pi^+ \pi^-)) = (-12.75 \pm 3.63_{\rm exp}\pm2.55_{\rm U-spin}\pm0.64_{\rm power} )\%,\notag\\
    &A_{CP}^{dir}(\Lambda_b^0 \to R( p \pi^-)R( K^+ K^-)) = (-65.93 \pm 20.06_{\rm exp}\pm13.90_{\rm U-spin}\pm3.30_{\rm power})\%,\notag\\
    &A_{CP}^{dir}(\Lambda_b^0 \to R( p K^-)R( K^+ K^-)) = (21.28 \pm 6.08_{\rm exp}\pm4.26_{\rm U-spin}\pm1.06_{\rm power})\%.\label{pre1}
\end{align}
Here the first error comes form the experimental data and the second error is estimated with symmetry breaking effect. The third error show the uncertainty form the power expansion. 
Due to the lack of experimental data, there are some CPV of b-baryon processes we cannot give the numerical prediction. The CPV of them can be expressed as
\begin{align}
&A_{CP}^{dir}(\Xi_b^0\to R(\Sigma^+ K^- K^+) K^-) =\frac{{(2.77\pm0.54)}\times 10^{-4}}{Br(\Xi_b^0\to \Sigma^+ K^-  K^- K^+)},\notag\\
&A_{CP}^{dir}(\Xi_b^0 \to R(\Sigma^+  K^+
K^-)\pi^- )=\frac{(-2.77\pm0.54)\times 10^{-4}}{Br(\Xi_b^0 \to \Sigma^+ \pi^- K^- K^+)},\notag\\
&A_{CP}^{dir}(\Xi_b^0\to R(\Sigma^+ K^-)R( K^+ K^-)) =\frac{{(2.72\pm0.73)}\times 10^{-4}}{Br(\Xi_b^0\to \Sigma^+ K^-  K^- K^+)},\notag\\
&A_{CP}^{dir}(\Xi_b^0\to R(\Sigma^+ K^-)R( \pi^+ \pi^-)) =\frac{{(2.72\pm0.73)}\times 10^{-4}}{Br(\Xi_b^0\to \Sigma^+ K^-  \pi^- \pi^+)},\notag\\
&A_{CP}^{dir}(\Xi_b^0\to R(\Sigma^+ \pi^-)R( \pi^+ \pi^-)) =\frac{{(-2.72\pm0.73)}\times 10^{-4}}{Br(\Xi_b^0\to \Sigma^+ \pi^-  \pi^- \pi^+)},\notag\\
&A_{CP}^{dir}(\Xi_b^0\to R(\Sigma^+ \pi^-)R( K^+ K^-)) =\frac{{(-2.72\pm0.73)}\times 10^{-4}}{Br(\Xi_b^0\to \Sigma^+ \pi^-  K^- K^+)}.\label{pre2}
\end{align}

In conclusion, our work provides a U-spin analysis on b-baryon four body decay CPV. Based on the recent measurement, we predict the CPV of $\Xi_b^0 \to \Sigma^+ \pi^- K^- K^+$ which is consistent with other work~\cite{He:2025msg}. To improve the predicted power of our work, we also perform a simple dynamic analysis and we can prove that for the specific decay $B_b^2\to R(B_1^2  M^2 M^{\bar 2})M^2$, only $c_{1q}+c_{3q}$ can contribute at leading power of $\sqrt{\frac{\Lambda_{QCD}}{m_b}}$, while for $B_b^2\to R(B_1^2  M^2)( M^{\bar 2}M^2)$, only $c_{1q}$ contributes at this leading power. Based on the analysis, we predict the direct CP-violating asymmetries $A_{CP}^{\text{dir}}$ as follows: $A_{CP}^{dir}(\Lambda_b^0 \to R( p \pi^- \pi^+ )\pi^-) = (-12.99 \pm 2.83\pm2.59\pm0.65)\%$, $A_{CP}^{dir}(\Lambda_b^0 \to R( p \pi^-)R( \pi^+ \pi^-)) = (-12.75 \pm 3.63\pm2.55\pm0.64)\%$, $A_{CP}^{dir}(\Lambda_b^0 \to R( p \pi^-)R( K^+ K^-)) = (-65.93 \pm 20.06\pm13.90\pm3.30)\%$ and $A_{CP}^{dir}(\Lambda_b^0 \to R( p K^-)R( K^+ K^-)) = (21.28 \pm 6.08\pm4.26\pm1.06)\%$. Considering $\Lambda_b$ can be effectively produced in LHCb, we strongly encourage a more precise experimental investigation of this channel. The other predictions are given in Eq.\eqref{pre1} and Eq.\eqref{pre2}.

%%%%%%%%%%%%%%%%%%%%%%%%
\section*{Acknowledgements}
%%%%%%%%%%%%%%%%%%%%%%%%
We thank Dr. Jin Sun, Mr. Chao-qi Zhang and Mr. Kai-wen Chen for useful discussion.
The work of Ruilin Zhu is supported by NSFC under grant No. 12322503 and No. 12075124, and by Natural Science Foundation of Jiangsu under Grant No. BK20211267.
The work of Zhi-Peng Xing is supported by NSFC under grant No.12375088 , No. 12335003 and No. 12405113.

$Note\;\;added$: While we were finishing our work, there are two preprint by Zhang, Wang [arXiv:2503.21885] , He, Liu, Tandean [arXiv:2503.24350]  and Shibasis Roy [arXiv:2504.10891] appeared, which also consider the U-spin for related decays. Compared with their works,  the U-spin symmetry combined the dynamic analysis are included.

%%%%%%%%%%%%%%%%%%%%%%%%%%%%%%%%%%%%%%%%%%%%%%%%%%%%%%

%%%%%%%%%%%%%%%%%%%%%%%%%%%%%%%%%%%%%%%%%%%%%%%%%%%%%%
\appendix
\section{The construction of U-spin matrix}\label{appendixA}
This section presents the construction of $\mathrm{U}$-spin representation matrices. The procedure involves:

\begin{enumerate}
    \item Deriving the $d$ and $s$ quark wave functions under $\mathrm{U}$-spin symmetry,
    \item Matching them with the corresponding SU(3)$_F$ particle states,
    \item Embedding the matched states into the representation matrices.
\end{enumerate}
For the meson system, the U-spin symmetry $2\otimes \bar 2 = 3 \oplus 1 $, the wave functions can be expressed as:
\begin{align}
  &\mathbf{3}:\left\{
    \begin{aligned}
      &U=1, U_3=+1: \quad \phi_3^{+1} = -d \bar s , \\
      &U=1, U_3=0: \quad \phi_3^0 = \frac{1}{\sqrt{2}} \left( d\bar d - s \bar s \right), \\
      &U=1, U_3=-1: \quad \phi_3^{-1} = s\bar d,
    \end{aligned}
  \right.\label{su2m}
  &\mathbf{1}:\quad U=0,U_3=0:\quad \phi_1^{0} =\frac{1}{\sqrt{2}} \left( d\bar d + s \bar s \right).
\end{align}
For the baryon system, the U-spin symmetry $2\otimes 2 = 3 \oplus 1 $, the wave functions can be expressed as:
\begin{align}
  &\mathbf{3}:\left\{
    \begin{aligned}
      &U=1, U_3=+1: \quad \phi_3^{+1} = dd , \\
      &U=1, U_3=0: \quad \phi_3^0 = \frac{1}{\sqrt{2}} \left( d s + s d \right), \\
      &U=1, U_3=-1: \quad \phi_3^{-1} = ss,
    \end{aligned}
  \right.
  &\mathbf{1}:\quad U=0,U_3=0:\quad \phi_1^{0} =\frac{1}{\sqrt{2}} \left( ds - s d \right).
\end{align}
U-spin symmetry $2 \otimes 2 \otimes 2 = 4 \oplus  2 \oplus  2^\prime $, the wave functions can be expressed as:
\begin{align}
      &\mathbf{4}:\left\{
    \begin{aligned}
      &U=3/2, U_3=+3/2: \quad \phi_4^{+3/2} = ddd, \\
      &U=3/2, U_3=+1/2: \quad \phi_4^{+1/2} = \frac{1}{\sqrt{3}} \left( dds + dsd +sdd \right), \\
      &U=3/2, U_3=-1/2: \quad \phi_4^{-1/2} = \frac{1}{\sqrt{3}} \left( ssd + dss +sds \right),\\
      &U=3/2, U_3=-3/2: \quad \phi_4^{-3/2} = sss,
    \end{aligned}
  \right.\\\notag\\
      &\mathbf{2}:\left\{
    \begin{aligned}
      &U=1/2, U_3=+1/2: \quad \phi_2^{+1/2} = \frac{1}{\sqrt{2}} \left( dsd - sdd\right), \\
      &U=1/2, U_3=-1/2: \quad \phi_2^{-1/2} = \frac{1}{\sqrt{2}} \left( dss - sds\right),
    \end{aligned}
  \right.\\\notag\\
        &\mathbf{2^\prime}:\left\{
    \begin{aligned}
      &U=1/2, U_3=+1/2: \quad \phi_{2^\prime}^{+1/2} = \frac{1}{\sqrt{6}} \left( dsd + sdd - 2 dds\right), \\
      &U=1/2, U_3=-1/2: \quad \phi_{2^\prime}^{-1/2} = \frac{1}{\sqrt{6}} \left( dss - sds + 2 ssd\right).
    \end{aligned}
  \right.
\end{align}
The meson and baryon wave functions in SU(3)$_F$ are constructed as follows:
\begin{align}
    &\pi^-=d\bar u,\quad K^-=s\bar u,\quad \pi^+=u\bar d,\quad  K^+=u\bar s,\notag\\
    &\pi^0 =  \frac{1}{\sqrt{2}}(u\bar u - d\bar d),\quad \eta_8 = \frac{1}{\sqrt{6}}(u\bar u + d\bar d - 2s\bar s),\quad K^0=d\bar s,\quad \bar K^0= s\bar d,\notag\\
    &\Xi_b^-=\frac{1}{\sqrt{2}}(ds-sd)b,\quad \Xi^0_b=\frac{1}{\sqrt{2}}(us-su)b,\quad \Lambda^0_b=\frac{1}{\sqrt{2}}(ud-du)b,\notag\\
    &p = \frac{1}{\sqrt{2}}[\frac{1}{\sqrt{6}}(2uud-udu-duu)\frac{1}{\sqrt{6}}(2\uparrow\uparrow\downarrow-\uparrow\downarrow\uparrow-\downarrow\uparrow\uparrow)+\frac{1}{\sqrt{2}}(udu-duu)\frac{1}{\sqrt{2}}(\uparrow\downarrow\uparrow-\downarrow\uparrow\uparrow)],\notag\\
    &\Sigma^+ = \frac{1}{\sqrt{2}}[\frac{1}{\sqrt{6}}(2uus-usu-suu)\frac{1}{\sqrt{6}}(2\uparrow\uparrow\downarrow-\uparrow\downarrow\uparrow-\downarrow\uparrow\uparrow)+\frac{1}{\sqrt{2}}(usu-suu)\frac{1}{\sqrt{2}}(\uparrow\downarrow\uparrow-\downarrow\uparrow\uparrow)],\notag\\
    &\Sigma^-=\frac{1}{\sqrt{2}}[\frac{1}{\sqrt{6}}(2dds-dsd-sdd)\frac{1}{\sqrt{6}}(2\uparrow\uparrow\downarrow-\uparrow\downarrow\uparrow-\downarrow\uparrow\uparrow)+\frac{1}{\sqrt{2}}(dsd-sdd)\frac{1}{\sqrt{2}}(\uparrow\downarrow\uparrow-\downarrow\uparrow\uparrow)], \notag\\
    & \Xi^-=\frac{1}{\sqrt{2}}[\frac{1}{\sqrt{6}}(sds+dss-2ssd))\frac{1}{\sqrt{6}}(2\uparrow\uparrow\downarrow-\uparrow\downarrow\uparrow-\downarrow\uparrow\uparrow)+\frac{1}{\sqrt{2}}(dss-sds)\frac{1}{\sqrt{2}}(\uparrow\downarrow\uparrow-\downarrow\uparrow\uparrow)],\notag\\
    &n=\frac{1}{\sqrt{2}}[\frac{1}{\sqrt{6}}(-2ddu+dud+udd)\frac{1}{\sqrt{6}}(2\uparrow\uparrow\downarrow-\uparrow\downarrow\uparrow-\downarrow\uparrow\uparrow)+\frac{1}{\sqrt{2}}(udd-dud)\frac{1}{\sqrt{2}}(\uparrow\downarrow\uparrow-\downarrow\uparrow\uparrow)], \notag\\
    &\Xi^0=\frac{1}{\sqrt{2}}[\frac{1}{\sqrt{6}}(usu+uss-ssu)\frac{1}{\sqrt{6}}(2\uparrow\uparrow\downarrow-\uparrow\downarrow\uparrow-\downarrow\uparrow\uparrow)+\frac{1}{\sqrt{2}}(uss-sus)\frac{1}{\sqrt{2}}(\uparrow\downarrow\uparrow-\downarrow\uparrow\uparrow)], \notag\\
    &\Sigma^0=\frac{1}{\sqrt{2}}[\frac{1}{\sqrt{12}}(2uds-usd-dsu+2dus-sud-sdu)\frac{1}{\sqrt{6}}(2\uparrow\uparrow\downarrow-\uparrow\downarrow\uparrow-\downarrow\uparrow\uparrow)\notag\\
    &\quad \quad+\frac{1}{2}(usd+dsu-sdu-sud)\frac{1}{\sqrt{2}}(\uparrow\downarrow\uparrow-\downarrow\uparrow\uparrow)], \notag\\
    &\Lambda^0=\frac{1}{\sqrt{2}}[\frac{1}{\sqrt{12}}(2uds-dsu-sud-2dus-sdu+usd)\frac{1}{\sqrt{2}}(\uparrow\downarrow\uparrow-\downarrow\uparrow\uparrow)\notag\\
    &\quad \quad+\frac{1}{2}(usd+sud-sdu-dsu)\frac{1}{\sqrt{6}}(2\uparrow\uparrow\downarrow-\uparrow\downarrow\uparrow-\downarrow\uparrow\uparrow)],\label{wavef}
\end{align}
where baryon octet wave functions must be symmetric under flavor-spin combinations to satisfy the Pauli principle.

Since the $\pi^-$, $\pi^+$, $K^+$, $K^-$, $\Xi^0_b$, $\Lambda^0_b$, $p$ and $\Sigma^+$ contain just a single down-type quark, they can be directly decomposed into U-spin doublets: $B_{b}^2=(\Lambda_b^0,\Xi_b^0),\;M^2=(\pi^-,K^-),\;M^{\bar 2}=(\pi^+,K^+),\;B_1^2=(p,\Sigma^+)$. For the meson triplet, by comparing the wave functions of $\pi^0$, $K^0$, $\bar{K}^0$, and $\eta_8$ under both $\mathrm{U}$-spin and $\mathrm{SU}(3)_F$ symmetries in Eq.\eqref{su2m} and  Eq.\eqref{wavef}, we can derive
\begin{align}
    \phi^{+1}_3=d\bar s= K^0,\phi^0_3=\frac{1}{\sqrt{2}}(d\bar d - s \bar s) = \frac{\sqrt{3}\eta_8}{2}-\frac{\pi^0}{2},\phi^{-1}_3=s\bar d = \bar K^0.
\end{align}
Then the U-spin representation of $(M^3)^i_j$ can be expressed as
\begin{align}
    (M^3)^i_j=\begin{pmatrix}
\phi_3^{0}&\sqrt{2}\phi_3^{+1} \\
\sqrt{2}\phi_3^{-1}&-\phi_3^{0} \\
\end{pmatrix}=\begin{pmatrix}
\frac{\sqrt{3}\eta_8}{2}-\frac{\pi^0}{2}&\sqrt{2}K^0 \\
\sqrt{2}\bar K^0&\frac{\pi^0}{2}-\frac{\sqrt{3}\eta_8}{2} \\
\end{pmatrix}.
\end{align}  

For the $\Sigma^-$ and $\Xi^-$ baryons, their quark compositions are $dds$ and $dss$, respectively. The functions under U-spin symmetry can be constructed as follows:
\begin{align}
\mathbf{4}:\left\{
\begin{aligned}
    &\phi_4(dds)=\frac{1}{\sqrt{3}} \left( dds + dsd +sdd \right)\frac{1}{\sqrt{3}} \left( \uparrow\uparrow\downarrow + \uparrow\downarrow\uparrow +\downarrow\uparrow\uparrow \right),\\
    &\phi_4(dss) = \frac{1}{\sqrt{3}} \left( ssd + dss +sds \right)\frac{1}{\sqrt{3}} \left( \uparrow\uparrow\downarrow + \uparrow\downarrow\uparrow +\downarrow\uparrow\uparrow \right).
\end{aligned}
     \right.
\end{align}
\begin{align}
\mathbf{2}:\left\{
\begin{aligned}
    &\phi_2(dds)=\frac{1}{\sqrt{2}}[\frac{1}{\sqrt{6}}(2dds-dsd-sdd)\frac{1}{\sqrt{6}}(2\uparrow\uparrow\downarrow-\uparrow\downarrow\uparrow-\downarrow\uparrow\uparrow)+\frac{1}{\sqrt{2}}(dsd-sdd)\frac{1}{\sqrt{2}}(\uparrow\downarrow\uparrow-\downarrow\uparrow\uparrow)],\\
    &\phi_2(dss) = \frac{1}{\sqrt{2}}[\frac{1}{\sqrt{6}}(sds+dss-2ssd))\frac{1}{\sqrt{6}}(2\uparrow\uparrow\downarrow-\uparrow\downarrow\uparrow-\downarrow\uparrow\uparrow)+\frac{1}{\sqrt{2}}(dss-sds)\frac{1}{\sqrt{2}}(\uparrow\downarrow\uparrow-\downarrow\uparrow\uparrow)].\label{sigxi}
\end{aligned}
     \right.
\end{align}
Comparing the wave functions of $\Sigma^-$ and $\Xi^-$ under both U-spin and SU(3)$_F$ symmetries in Eq.\eqref{sigxi} and Eq.\eqref{wavef}, the U-spin representation of $B^2_2$ can expressed as
\begin{align}
\phi_2(dds)=\Sigma^-,\phi_2(dss)=\Xi^-,B^2_2=(\phi_2(dds),\phi_2(dss))=(\Sigma^-,\Xi^-).
\end{align}
The derivation of matrix $B^3$ and $B^1$ proceeds through the construction of U-spin wave functions. Initially, the u quark and spin are introduced without explicit symmetry restrictions. The symmetric wave functions are then generated through the application of permutation operation as
\begin{align}
  &\mathbf{3}:\left\{
    \begin{aligned}
      &U=1, U_3=+1: \quad \phi_3^{+1} = -\frac{1}{\sqrt{3}}[dd\frac{1}{\sqrt{2}}(\uparrow\downarrow+\downarrow\uparrow),u\uparrow]_p+\frac{\sqrt{2}}{\sqrt{3}}[dd\uparrow\uparrow,u\downarrow]_p, \\
      &U=1, U_3=0: \quad \phi_3^0 = -\frac{1}{\sqrt{3}}[\frac{1}{\sqrt{2}}(ds+sd)\frac{1}{\sqrt{2}}(\uparrow\downarrow+\downarrow\uparrow),u\uparrow]_p+\frac{\sqrt{2}}{\sqrt{3}}[\frac{1}{\sqrt{2}}(ds+sd)\uparrow\uparrow,u\downarrow]_p, \\
      &U=1, U_3=-1: \quad \phi_3^{-1} = -\frac{1}{\sqrt{3}}[ss\frac{1}{\sqrt{2}}(\uparrow\downarrow+\downarrow\uparrow),u\uparrow]_p+\frac{\sqrt{2}}{\sqrt{3}}[ss\uparrow\uparrow,u\downarrow]_p,\label{tritri}
    \end{aligned}
  \right.\\\notag\\
  &\mathbf{1}:\quad U=0,U_3=0:\quad \phi_1^{0} =-\frac{1}{\sqrt{3}}[\frac{1}{\sqrt{2}}(ds-sd)\frac{1}{\sqrt{2}}(\uparrow\downarrow+\downarrow\uparrow),u\uparrow]_p+\frac{\sqrt{2}}{\sqrt{3}}[\frac{1}{\sqrt{2}}(ds-sd)\uparrow\uparrow,u\downarrow]_p,\label{trione}
\end{align}
where $[ds,u]_p$ is the permutation operation which is $[ds,u]_p=\frac{1}{\sqrt{3}}(uds+dus+dsu)$. Comparing the wave function under U-spin and SU(3)$_F$ in Eq.\eqref{trione}, Eq.\eqref{tritri} and Eq.\eqref{wavef}, we can derive
\begin{align}
    \phi_3^{+1}=n,\quad\phi_3^{0}=-\frac{1}{2}\Sigma^0-\frac{\sqrt{3}}{2}\Lambda^0,\quad\phi_3^{-1}=\Xi^0,\quad\phi_1^{0}=-\frac{1}{2}\Lambda^0+\frac{\sqrt{3}}{2}\Sigma^0.
\end{align}
The SU(2)$_F$representation of baryon triplet and singlet can be expressed as
\begin{align}
B^3=\begin{pmatrix}
\phi_3^{0}&\sqrt{2}\phi_3^{+1} \\
\sqrt{2}\phi_3^{-1}&-\phi_3^{0} \\
\end{pmatrix}=\begin{pmatrix}
-\frac{\Sigma^0}{2\sqrt{2}}-\frac{\sqrt{3}\Lambda^0}{2}& \sqrt{2}n \\
\sqrt{2}\Xi^0&\frac{\Sigma^0}{2}+\frac{\sqrt{3}\Lambda^0}{2\sqrt{2}}\end{pmatrix},\quad B^1=\phi_1^{0}=-\frac{1}{2}\Lambda^0+\frac{\sqrt{3}}{2}\Sigma^0.
\end{align}

\end{document}